\documentclass[aps,pra,twocolumn,groupedaddress]{revtex4-2}

\usepackage{amsfonts}
\usepackage{amssymb}
\usepackage{amsmath}
\usepackage[english]{babel}
\usepackage{tumcolors}
\usepackage{tummath}
\usepackage{siunitx}
\usepackage{tikz}
\usepackage[straightvoltages, europeanresistors]{circuitikz}
\usepackage{pgfplots}
\usepackage{pgfplotstable}
\usepackage{environ}
\usepackage[hidelinks]{hyperref}
\usepackage[autostyle, english=american]{csquotes}
\MakeOuterQuote{"}

\newcommand{\figref}[2][{}]{\hyperref[#2]{\figurename~\ref{#2}#1}}
\newcommand{\tabref}[2][{}]{\hyperref[#2]{\tablename~\ref{#2}#1}}
\renewcommand{\imu}{\i}

\usetikzlibrary{arrows, decorations.markings}
\usepgfplotslibrary{units,fillbetween}
\pgfplotsset{compat=newest}

\pgfplotsset{
  layers/axis lines on top/.define layer set={
      axis background,
      axis grid,
      axis ticks,
      axis tick labels,
      pre main,
      main,
      axis lines,
      axis descriptions,
      axis foreground,
    }{/pgfplots/layers/standard},
}

\begin{document}

\title{Circuit~quantum~electrodynamic~model~of~dissipative-dispersive\\
  Josephson~traveling-wave~parametric~amplifiers}

\author{Yongjie \surname{Yuan}}
\author{Michael \surname{Haider}}
\email[]{michael.haider@tum.de}
\author{Johannes A. \surname{Russer}}
\author{Peter \surname{Russer}}
\author{Christian \surname{Jirauschek}}
\affiliation{TUM School of Computation, Information and Technology, Technical
  University of Munich, Hans-Piloty-Str.\ 1, 85748 Garching, Germany}

\date{\today}

\begin{abstract}
  We present a quantum mechanical model for a four-wave mixing Josephson traveling-wave parametric amplifier including substrate losses and associated thermal fluctuations. Under the assumption of a strong undepleted classical pump tone, we derive an analytic solution for the bosonic annihilation operator of the weak signal photon field using temporal equations of motion in a reference timeframe, including chromatic dispersion. From this result, we can predict the asymmetric gain spectrum of a Josephson traveling-wave parametric amplifier due to non-zero substrate losses. We also predict the equivalent added input noise including quantum fluctuations as well as thermal noise contributions. Our results are in excellent agreement with recently published experimental data.
\end{abstract}

\maketitle

\section{\label{sec-introduction}Introduction}

In conventional low-noise microwave amplifiers, amplification is achieved by
modulating the channel of a high-electron-mobility transistor (HEMT). A
characteristic measure for the amplifier's performance is the noise
temperature, which is defined as the equivalent temperature of a resistor
that would produce the same level of Johnson-Nyquist
noise~\cite{johnson_thermal_1928,nyquist_thermal_1928}. The added noise power
depends on the channel resistance and easily exceeds the energy of a few tens
of microwave photons, even when the circuit is cooled down to cryogenic
temperatures~\cite{abdo_josephson_2011}. In superconducting quantum computing
where the qubit state is probed by ultra-low-power microwave signals,
however, quantum-limited~\cite{caves_quantum_1982} noise performance is key
for high-fidelity single-shot readouts of quantum
information~\cite{reed_high-fidelity_2010}. This and the limited cooling
power budget in the lowest temperature stage of dilution refrigerators render
the use of traditional solid-state amplifiers impossible in the context of
low-power dispersive qubit readout.

For the detection and amplification of single-photon-level microwave signals,
a different type of low-noise amplifier is required where the gain does not
originate from modulating a dissipative channel. Parametric amplification is
accomplished by nonlinear mixing of the input signal with a strong coherent
pump field. The wave-mixing interaction is non-dissipative and thus achieves
superior noise performance~\cite{clerk_introduction_2010}. Superconducting
parametric amplifiers based on the nonlinear kinetic inductance of Josephson
junctions~\cite{josephson_possible_1962} approach the quantum noise
limit~\cite{caves_quantum_1982} with very little power dissipation. Hence,
they are used as first-stage amplifiers in the readout of superconducting
qubits~\cite{johansson_readout_2006,wendin_quantum_2007,mallet_single-shot_2009}.
In recent years, research on parametric amplifiers gained a lot of momentum
due to the growing interest in superconducting quantum computing. The first
experimental evidence for the feasibility of parametric amplification with
Josephson junctions was already obtained in
1967~\cite{zimmer_parametric_1967}. The first theoretical study was given
in~\cite{russer_parametric_1969}. General energy relations for frequency
conversion in nonlinear reactances have been found
in~\cite{russer_general_1971}, which are needed to describe DC-pumped
parametric amplification~\cite{russer_nanoelectronic_2011}. An exhaustive
review of different microwave parametric amplifier designs in the context of
quantum information experiments is given in~\cite{gu_microwave_2017}. In a
typical architecture, a single Josephson junction is coupled to a resonator
in order to increase the gain of the amplifier by increasing the interaction
time of the signal and pump modes. The microwave resonator, however, limits
the bandwidth of the amplifier to the resonator bandwidth. The amplifier
resonator is operated in reflection mode, where a bulky microwave circulator
needs to be used to separate the respective input and output
waves~\cite{mutus_strong_2014}. Both limitations, the limited bandwidth as
well as the need for a circulator, can be overcome by using a traveling-wave
type architecture as proposed by~\cite{macklin_near-quantum-limited_2015},
where Josephson junctions are periodically embedded in a microwave
transmission line. Josephson traveling-wave parametric amplifiers (JTWPAs)
achieve large gain by increasing the interaction time of the signal and pump
modes through a large propagation
distance~\cite{russer_dc-pumped_1977,russer_circuit_1979}. The interaction
along the nonlinear transmission line shows a strong phase sensitivity. Thus,
optimum parametric gain can only be achieved if the amplification process is
phase-matched by careful dispersion
engineering~\cite{obrien_resonant_2014,planat_photonic-crystal_2020,ranadive_kerr_2022}.

In~\cite{yaakobi_parametric_2013}, the Josephson-embedded nonlinear
transmission line is described by a classical nonlinear wave equation. The
operation principle was modeled using coupled-mode equations (CMEs), allowing
to straightforwardly calculate the amplifier's gain spectrum. Single-photon
applications at ultra-low temperatures, however, necessitate a quantum
mechanical treatment of the device. Quantum mechanical models for Josephson
circuits can be obtained in the framework of quantum circuit
theory~\cite{haus_steady-state_1970,yurke_quantum_1984,haus_quantum_1987},
where the term mesoscopic physics has been coined for systems which contain a
large number of electons and yet show distinct quantum
features~\cite{vool_introduction_2017}.

Quantum models for lumped-element Josephson parametric amplifiers (JPAs) have
been investigated
in~\cite{kaertner_generation_1990,russer_nanoelectronic_2011,armour_josephson_2015}.
Dissipation and thermal noise in the JPA were introduced phenomenologically
in~\cite{kaiser_quantum_2017} by coupling the inner degrees of freedom to a
bath. A Hamiltonian description of a JTWPA using continuous-mode operators
was derived in~\cite{grimsmo_squeezing_2017}. A similar discrete-mode
mesoscopic Hamiltonian was presented in~\cite{reep_mesoscopic_2019}, which
serves as a starting point for the investigations of the present paper. For
the derivation of the CMEs
in~\cite{yaakobi_parametric_2013,obrien_resonant_2014} and the quantum
descriptions in~\cite{grimsmo_squeezing_2017,reep_mesoscopic_2019}
non-dissipative parametric amplification was assumed, i.e.\ all losses were
neglected.

In this work, we extend the Hamiltonian framework
from~\cite{grimsmo_squeezing_2017,reep_mesoscopic_2019} in order to treat
noise and dissipation, mainly arising from substrate losses along the
transmission line since the transmission line conductor itself is in a
superconducting state. Noise and dissipation are included in the Hamiltonian
by phenomenologically coupling the signal and idler modes to a bath,
consisting of an infinite number of quantum harmonic
oscillators~\cite{sargent_laser_1974}. The coupling constants depend on the
substrate material and can be represented by different coupling
models~\cite{haider_including_2021}. Having a dissipative model for a
Josephson traveling-wave parametric amplifier at hand enables studying the
frequency-dependent attenuation and noise performance close to the quantum
limit. Fluctuations and dissipation in quantum traveling-wave parametric
amplifiers were studied in~\cite{houde_loss_2019} using input-output theory
as well as a distributed loss model. In the following, we present an analytic
solution for the photon field annihilation operator in a Josephson-embedded
transmission line including substrate losses and additional thermal noise.
From there we calculate expressions for the amplifier gain and the added
input noise which both match experimental
results~\cite{simbierowicz_characterizing_2021}. We predict an added input
noise equivalent to approximately \num{1.3} excess photons, which is in
excellent agreement with recent experimental
observations~\cite{simbierowicz_characterizing_2021}.

First, we introduce a circuit model of a dissipative-dispersive Josephson
junction-embedded transmission line in Section~\ref{sec-quantum-treatment}.
We then discuss the traveling-wave mode quantization in
Section~\ref{sec-mode-quantization} and introduce the JTWPA dispersion
relation along with a reference timeframe in Section~\ref{sec-time-frame}. In
Section~\ref{sec-hamiltonian} we introduce our four-wave mixing Hamiltonian.
Afterwards, the Heisenberg equations of motion are derived within the moving
reference timeframe in Section~\ref{sec-heisenberg}. The resulting equations
of motion for the photon field are then solved in order to get an analytic
expression for the signal mode annihilation operator under a strong classical
pump approximation in Section~\ref{sec-analytic-solution}. Next, we present
analytic results for the gain profile and the temporal dynamics of an
exemplary JTWPA structure from the
literature~\cite{obrien_resonant_2014,reep_mesoscopic_2019} including
substrate losses in Section~\ref{sec-gain-evolution}. In
Section~\ref{sec-noise} we derive an analytic expression for the number of
added noise photons due to thermal fluctuations. We conclude the section with
a discussion on the equivalent added input noise for an experimental
structure~\cite{simbierowicz_characterizing_2021}, where our theoretical
predictions are found to be in excellent agreement with their observations.

\section{\label{sec-quantum-treatment}Quantum mechanical treatment of a dissipative JTWPA}

\begin{figure}
  \centering
  \begin{tikzpicture}
    \ctikzset{bipoles/length=1cm}
    \draw (0,0) to[short, o-] (2,0) to[short] (2,0.5) to[C, l=$C_\mathrm{J}$] (3.5,0.5) to[short] (3.5,0) to[short, -o] (7,0)
      (2,0) to[short, *-] (2,-0.5) to[barrier, /tikz/circuitikz/bipoles/length=2.5cm] (3.5,-0.5) to[short, -*] (3.5,0);
    \node[below=2.5mm] at (2.75,-0.5) {$L_\mathrm{J}$};
    \draw (1,0) to[C, *-*, l=$C'\Delta{z}$] (1,-2.5);
    \draw (0,-2.5) to[short, o-o] (7,-2.5);
    \draw (0,0) to[open, v=$v_n$] (0,-2.5);
    \draw (7,0) to[open, v^=$v_{n+1}$] (7,-2.5);
    \draw[color=TUMBlue] (4.5,0) to[C, *-, l_=$C_\mathrm{c}$] (4.5,-1.25) to[short] (5,-1.25) to[L, -*, l_=$L_\mathrm{r}$] (5,-2.5)
      (4.5,-1.25) to[short, *-] (4,-1.25) to[C, -*, l_=$C_\mathrm{r}$] (4,-2.5);
    \draw[color=TUMOrange] (5.75,0) to[R, *-*] (5.75,-2.5)
      (5.75, -0.5) to[short, *-] (6.5,-0.5) to[isource] (6.5,-2) to[short, -*] (5.75,-2);
    \draw[latex-latex] (0,-3) -- (7,-3) node [midway, below] {$\Delta{z}$};
  \end{tikzpicture}
  \caption{Unit cell of a JTWPA. The resonant phase matching circuit and a resistive representation of the bath are highlighted in blue and orange color, respectively. The bath adds noise to the system via the current source, and dissipates energy through the resistor.}
  \label{fig-unit-cell}
\end{figure}
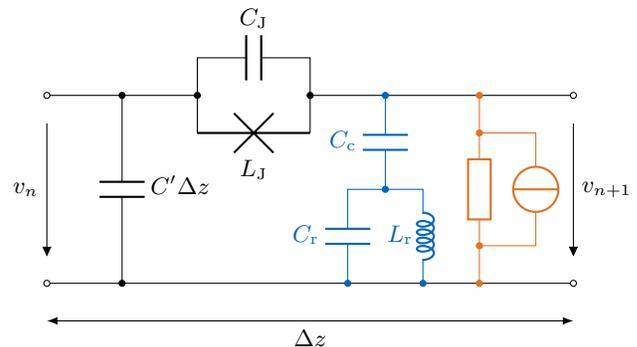

Consider a superconducting transmission line with periodically embedded
identical Josephson junction loadings. The distance between the Josephson
nonlinearities is small compared to the wavelength under consideration, which
allows for a continuum treatment of the Josephson embedded transmission
line~\cite{yaakobi_parametric_2013}. A circuit representation of the unit
cell of such a stucture is given in~\figref{fig-unit-cell}. As the
transmission line itself is in a superconducting state, losses and thermal
fluctuations only occur due to substrate imperfections, which are included in
terms of the resistor and the associated noise current source in the circuit
model in~\figref{fig-unit-cell}, highlighted in orange color. In order to
derive a quantum model of the continuous dissipative nonlinear transmission
line, the resistor and the associated noise current source are modeled in
terms of a distributed Markovian heat-bath, consisting of an infinite number
of harmonic oscillators with thermal initial occupations. It has been shown
in~\cite{obrien_resonant_2014} that the gain and bandwith of Josephson
traveling-wave parametric amplifiers can be significantly improved by
dispersion engineering using resonant phase-matching (RPM). Resonant
phase-matching can be included using periodically embedded LC resonators,
which are capacitively coupled to the transmission line. The resonators
tailor the dispersion relation such that the total phase mismatch along the
transmission line remains small over a large bandwidth.
In~\figref{fig-unit-cell}, and throughout the rest of this paper,
$C^{\prime}$ is the ground capacitance per unit length of the transmission
line, $L_{\mathrm{J},0}$ and $C_{\mathrm{J}}$ are the linear inductance and
intrinsic capacitance of the Josephson junctions, $\Delta{z}$ is the length
of a single unit cell, $C_\mathrm{c}$ is the RPM coupling capacitance, and
$C_\mathrm{r}$ and $L_\mathrm{r}$ are the RPM resonator capacitance and
inductance. Note that the transmission line inductance has been neglected, as
it can be easily included into the linear Josephson inductance.

\section{\label{sec-mode-quantization}Quantization of discrete and continuous modes}

A discrete-mode mesoscopic Hamiltonian for a non-dissipative
Josephson-embedded transmission line has been derived
in~\cite{reep_mesoscopic_2019}. For the quantum mechanical model in this work
we take a similar approach, where we however consider a continuous mode
spectrum in order to derive a consistent noise model in terms of a noise
spectral density~\cite{davenport_introduction_1987,koch_quantum-noise_1980}.
We use canonical quantization of right-moving traveling-wave
modes~\cite{vool_introduction_2017}, where the magnetic flux through the
Josephson element $\Delta{\Phi_{\mathrm{J}}}$ and the conjugate charge $Q$ in
the ground capacitance per unit length $C^{\prime}$ are used as canonical
pair of variables, defined over a continuous spatial argument $z$. Assuming a
monochromatic right-traveling wave propagating through a transmission line,
we can describe the voltage operator by
\begin{equation}
  \function{\op{V}}{z,t}=\function{\op{c}}{z}\,\e^{\imu\function{k}{\omega}z-\imu\omega t}+\mathrm{H.c.}\, ,
\end{equation}
where $\function{\op{c}}{z}$ is the amplitude operator, $\omega$ is the
angular frequency of the propagating wave, $\function{k}{\omega}$ is the
associated wave number, and $\mathrm{H.c.}$ denotes the Hermitian conjugate.

A right-traveling wave where the voltage is defined over a continuous
frequency spectrum, on the other hand, can be decomposed into monochromatic
plane wave contributions in terms of the inverse Fourier transform
\begin{equation}
  \function{\op{V}}{z,t}=\frac{1}{\sqrt{2\pi}}\int\limits_{-\infty}^{\infty}\op{c}_{\omega^\prime}\,\e^{\imu\function{k}{\omega^{\prime}}z-\imu\omega^{\prime}t}\diff{\omega^{\prime}}\, ,
\end{equation}
where the spectral Fourier coefficient operator $\op{c}_{\omega^{\prime}}$
can be obtained by means of the Fourier transform of the voltage operator
\begin{equation}
  \function{\op{c}_{\omega^{\prime}}}{z}=\frac{1}{\sqrt{2\pi}}\int\limits_{-\infty}^{\infty}\function{\op{V}}{z,t}\,\e^{-\imu\function{k}{\omega^{\prime}}z+\imu\omega^{\prime}t}\diff{t}\, .
\end{equation}
Hence, monochromatic right-traveling wave amplitude operators can be
represented by the spectral coefficient operators
\begin{align}
  \op{c}_{\omega^{\prime}}= & \frac{1}{\sqrt{2\pi}}\int\limits_{-\infty}^{\infty}\function{\op{c}}{z}\,\e^{\imu\left[\function{k}{\omega}-\function{k}{\omega^{\prime}}\right]z}\,\e^{-\imu\left(\omega-\omega^{\prime}\right)t}\diff{t}\nonumber \\
  =                         & \sqrt{2\pi}\function{\op{c}}{z}\function{\delta}{\omega^{\prime}-\omega}\, .\label{eqn-continuous-discrete-translation}
\end{align}
In the following, relation~\eqref{eqn-continuous-discrete-translation} will
be used to translate continuous frequency amplitude operators to their
monochromatic limits.

\section{\label{sec-time-frame}Dispersion and reference timeframe}

Traveling-wave amplitudes inside a dispersionless nonlinear transmission line
exhibit space-time translation invariance~\cite{roy_introduction_2016}, i.e.\
\begin{equation}
  \function{A}{z,t}=\function*{A}{0,t-\frac{z}{v}}=\function{A}{z-vt,0}\, ,
\end{equation}
with the spatial argument $z$ and the frequency independent propagation
velocity $v$. Thus, the evolution of a corresponding wave amplitude operator
$\function{\op{A}}{z,t}$ can be described by either a spatial or a temporal
dependence
\begin{equation}
  \function{\op{A}}{z,t}\to\function{\op{A}}{z}\Longleftrightarrow\function*{\op{A}}{t=\frac{z}{v}}\, ,
\end{equation}
where again only right-propagating waves are taken into account. In case of a
dispersive transmission line, the phase velocity
$\function{v_{\mathrm{ph}}}{\omega}=\omega/\function{k}{\omega}$ is
frequency-dependent, where $\function{k}{\omega}$ represents the dispersion
relation. As each mode now travels at a different velocity, distinct
frequency components arrive at a certain location $x$ at different times. In
other words, when looking at a certain location $x$ along the transmission
line, each mode travels within its own frequency-dependent timeframe. The
corresponding timeframes $\function{t}{\omega}$ are considered to be
frequency-dependent functions which are given by
$\function{t}{\omega}=z/\function{v_{\mathrm{ph}}}{\omega}$. Therefore, the
system operators can be described similar to the dispersionless case, by a
frequency-dependent time argument
\begin{equation}
  \function{\op{a}_{\omega}}{z}\to\function*{\op{a}_{\omega}}{\function{t}{\omega}=\frac{z}{\function{v_{\mathrm{ph}}}{\omega}}}\, .
\end{equation}
This way, the spatial dependence of the system annihilation operator
$\op{a}_{\omega}$ can be expressed by a temporal dependence which permits the
use of Heisenberg's equations of motion for calculating the temporal
evolution of the system. However, since the Heisenberg equations for
different modes are given by means of a frequency-dependent, and thus
mode-dependent timeframe $\function{t}{\omega}$, it is difficult to find
analytic solutions. In order to circumvent this problem, we introduce a
frequency-independent reference velocity
$v_{\mathrm{r}}=\sqrt{{\Delta{z}}/{L_{\mathrm{J},0}C^\prime}}$ with a
corresponding reference timeframe $t_{\mathrm{r}}={z}/{v_{\mathrm{r}}}$ which
is associated with a wave propagating in a dispersionless ideal transmission
line with capacitance $C^\prime$ and inductance $L_{\mathrm{J},0}$ per unit
length, respectively. Accordingly, we can define the translation between the
reference and phase timeframes by ${\partial{\function{t}{\omega}}}/{\partial{t_{\mathrm{r}}}}=\sqrt{\function{\Lambda}{\omega}}$. The dimensionless
dispersion factor $\function{\Lambda}{\omega}$ is given by
\begin{equation}
  \function{\Lambda}{\omega}=\frac{1+\sqrt{1+\frac{1}{\omega^2R^2{C^\prime}^2\Delta{z}^2}}}{2\left(1-\omega^2L_{\mathrm{J},0}C_{\mathrm{J}}\right)}\approx\frac{1}{1-\omega^2L_{\mathrm{J},0}C_{\mathrm{J}}}\, ,
  \label{eqn-dispersion-relation}
\end{equation}
and may include dispersion due to the intrinsic substrate resistance $R$, the
shunt capacitance $C^{\prime}$, as well as the Josephson capacitance
$C_{\mathrm{J}}$. However, the contributions of $R$ and $C^{\prime}$ can
usually be neglected. Note that, different from~\cite{reep_mesoscopic_2019},
the nonlinearity of the Josephson inductance is not explicitly taken into
account for deriving the dispersion factor $\function{\Lambda}{\omega}$,
where we only use the linear Josephson inductance $L_{\mathrm{J},0}$.
Ignoring substrate losses, however, \eqref{eqn-dispersion-relation} exactly
matches the dispersion relation in~\cite{obrien_resonant_2014}.

\section{\label{sec-hamiltonian}Four-wave mixing Hamiltonian for a JTWPA}

In order to provide a consistent investigation of thermal noise, we construct
the system Hamiltonian in terms of operators with a continuous mode spectrum.
Hence, we use~\cite{vool_introduction_2017}
\begin{equation*}
  \op{V}=\frac{1}{\sqrt{2\pi}}\int\limits_{0}^{\infty}\sqrt{\frac{\hbar\omega}{2C^{\prime}\function{v_{\mathrm{ph}}}{\omega}}}\left[\op{a}_\omega\e^{\imu\function{k}{\omega}z-\imu\omega t}+\mathrm{H.c.}\right]\diff{\omega}\, ,
\end{equation*}
where $C'$ is the ground capacitance per unit length and
$\function{v_{\mathrm{ph}}}{\omega}$ represents the frequency-dependent phase
velocity. Accordingly, the magnetic flux through the Josephson element can be
expressed by the operator
\begin{equation}
  \begin{split}
    \Delta{\op{\Phi}_{\mathrm{J}}}= & \frac{1}{\sqrt{2\pi}}\int\limits_{0}^{\infty}\frac{\function{k}{\omega}\Delta{z}}{\omega}\times\\
    & \times\sqrt{\frac{\hbar\omega}{2C'\function{v_{\mathrm{ph}}}{\omega}}}\left[\op{a}_\omega\,\e^{\imu\function{k}{\omega}z-\imu\omega t}+\mathrm{H.c.}\right]\diff{\omega}\, .
  \end{split}
  \label{eqn-josephson-flux}
\end{equation}
The Hamiltonian of a lossless JTWPA can then be obtained by integrating the
Hamiltonian density, i.e.\ the energy stored in the ground capacitors and in
the linear and nonlinear inductances of the Josephson junctions per unit
length, over the entire device length $x$~\cite{reep_mesoscopic_2019}
\begin{equation}
  \begin{split}
    \op{H}_{\mathrm{JTWPA}}&=\frac{1}{2\Delta{z}^2}\int\limits_{0}^{x}\left\lbrace\left[\frac{\Delta{z}}{L_{\mathrm{J,0}}}\Delta{\op{\Phi}_{\mathrm{J}}}-\frac{\Delta{z}}{12L_{\mathrm{J},0}\varphi_0^2}\Delta{\op{\Phi}_{\mathrm{J}}}^3\vphantom{\pdd[2]{\Delta{\op{\Phi}_{\mathrm{J}}}}{t}}\right.\right. \\
    & \qquad\left.\left.+C_{\mathrm{J}}\Delta{z}\pdd[2]{\Delta{\op{\Phi}_{\mathrm{J}}}}{t}\right]\Delta{\op{\Phi}_{\mathrm{J}}}+\frac{1}{C'}\op{Q}^2\right\rbrace\diff{z}\, ,
  \end{split}
  \label{eqn-jtwpa-hamiltonian}
\end{equation}
where $\varphi_0 = \hbar/\left(2e\right)$ is the reduced magnetic flux
quantum, $\hbar$ is the reduced Planck's quantum of action and $e$ is the
elementary charge. The quadratic terms in~\eqref{eqn-jtwpa-hamiltonian}
contribute to the linear part of the system Hamiltonian, while the
fourth-order term describes the nonlinearity which is responsible for the
four-wave-mixing amplification process. The continuous mode Hamiltonian
derived in~\eqref{eqn-jtwpa-hamiltonian} agrees with the Hamiltonian in
equation (9) of~\cite{grimsmo_squeezing_2017}.

The input mode spectrum can be decomposed into weak input signal components
$\op{a}_\omega$, together with a strong pump tone
$\op{a}_{\Omega_{\scriptsize\mathrm{p}}}$ at a pump frequency of
$\Omega_\mathrm{p}$. Inserting this decomposition into the Josephson flux
operator $\Delta{\Phi_{\mathrm{J}}}$ ~\eqref{eqn-josephson-flux} and then
inserting the result into the Hamiltonian~\eqref{eqn-jtwpa-hamiltonian}, we
obtain the lossless system Hamiltonian in terms of the continuous mode
operators $\op{a}_\omega$ and $\op{a}_{\Omega_{\scriptsize\mathrm{p}}}$
\begin{widetext}
  \begin{equation}
    \begin{split}
      \op{H}_{\mathrm{JTWPA}}= & \int\limits_{0}^{\infty}\hbar\Omega_{\mathrm{p}}\op{a}^\dagger_{\Omega_{\scriptsize\mathrm{p}}}\op{a}_{\Omega_{\scriptsize\mathrm{p}}}\diff{\Omega_{\mathrm{p}}}+\int\limits_{0}^{\infty}\hbar\omega\op{a}^\dagger_{\omega}\op{a}_{\omega}\diff{\omega}\\
      & -\frac{\hbar^2\Delta{z}^3}{64\pi^2 {C^{\prime}}^2L_{\mathrm{J},0}^3I_{\mathrm{c}}^2}\int\limits_{0}^{\infty}\!\int\limits_{0}^{\infty}\!\int\limits_{0}^{\infty}\!\int\limits_{0}^{\infty}\frac{\function{k}{\tilde{\Omega}_{\mathrm{p}}}\function{k}{\tilde{\Omega}^\prime_{\mathrm{p}}}\function{k}{\Omega_{\mathrm{p}}}\function{k}{\Omega^\prime_{\mathrm{p}}}\,\e^{-\imu\left(\tilde{\Omega}_\mathrm{p}^\prime-\tilde{\Omega}_\mathrm{p}+\Omega_\mathrm{p}^\prime-\Omega_\mathrm{p}\right)t}}{\sqrt{\tilde{\Omega}_{\mathrm{p}}\tilde{\Omega}^\prime_{\mathrm{p}}\Omega_{\mathrm{p}}\Omega_{\mathrm{p}}^{\prime}\function{v_{\mathrm{ph}}}{\tilde{\Omega}_{\mathrm{p}}}\function{v_{\mathrm{ph}}}{\tilde{\Omega}^\prime_{\mathrm{p}}}\function{v_{\mathrm{ph}}}{\Omega_{\mathrm{p}}}\function{v_{\mathrm{ph}}}{\Omega_{\mathrm{p}}^{\prime}}}}\op{a}^\dagger_{\tilde{\Omega}_{\scriptsize\mathrm{p}}}\op{a}_{\tilde{\Omega}^{\prime}_{\scriptsize\mathrm{p}}}\op{a}^\dagger_{\Omega_{\scriptsize\mathrm{p}}}\op{a}_{\Omega^\prime_{\scriptsize\mathrm{p}}}\times\\
      & \times\int\limits_{0}^{x}\e^{\imu\left[\function{k}{\tilde{\Omega}^\prime_{\mathrm{p}}}-\function{k}{\tilde{\Omega}_{\mathrm{p}}}+\function{k}{\Omega^\prime_{\mathrm{p}}}-\function{k}{\Omega_{\mathrm{p}}}\right]z}\,\diff{z}\diff{\tilde{\Omega}_{\mathrm{p}}}\diff{\tilde{\Omega}^{\prime}_{\mathrm{p}}}\diff{\Omega_{\mathrm{p}}}\diff{\Omega^{\prime}_{\mathrm{p}}}\\
      & -\frac{\hbar^2\Delta z^3}{16\pi^2{C^{\prime}}^2L^3_{\mathrm{J},0}I^2_{\mathrm{c}}}\int\limits_{0}^{\infty}\!\int\limits_{0}^{\infty}\!\int\limits_{0}^{\infty}\!\int\limits_{0}^{\infty}\frac{\function{k}{\omega}\function{k}{\omega^\prime}\function{k}{\Omega_{\mathrm{p}}}\function{k}{\Omega^\prime_{\mathrm{p}}}\,\e^{-\imu\left(\omega^\prime-\omega+\Omega_\mathrm{p}^\prime-\Omega_\mathrm{p}\right)t}}{\sqrt{\omega\omega^{\prime}\Omega_{\mathrm{p}}\Omega_{\mathrm{p}}^{\prime}\function{v_{\mathrm{ph}}}{\omega}\function{v_{\mathrm{ph}}}{\omega^{\prime}}\function{v_{\mathrm{ph}}}{\Omega_{\mathrm{p}}}\function{v_{\mathrm{ph}}}{\Omega_{\mathrm{p}}^{\prime}}}}\op{a}^\dagger_{\omega}\op{a}_{\omega^\prime}\op{a}^\dagger_{\Omega_{\scriptsize\mathrm{p}}}\op{a}_{\Omega^\prime_{\scriptsize\mathrm{p}}}\times\\
      & \times\int\limits_{0}^{x}\e^{\imu\left[\function{k}{\omega^\prime}-\function{k}{\omega}+\function{k}{\Omega^\prime_{\mathrm{p}}}-\function{k}{\Omega_{\mathrm{p}}}\right]z}\,\diff{z}\diff{\omega}\diff{\omega^{\prime}}\diff{\Omega_{\mathrm{p}}}\diff{\Omega^{\prime}_{\mathrm{p}}}\\
      & -\frac{\hbar^2\Delta z^3}{32\pi^2{C^{\prime}}^2L^3_{\mathrm{J},0}I^2_{\mathrm{c}}}\int\limits_{0}^{\infty}\!\int\limits_{0}^{\infty}\!\int\limits_{0}^{\infty}\!\int\limits_{0}^{\infty}\left[\frac{\function{k}{\omega}\function{k}{\omega^{\prime}}\function{k}{\Omega_{\mathrm{p}}}\function{k}{\Omega^{\prime}_{\mathrm{p}}}\,\e^{-\imu\left(-\omega-\omega^\prime+\Omega_\mathrm{p}+\Omega_\mathrm{p}^\prime\right)t}}{\sqrt{\omega\omega^{\prime}\Omega_{\mathrm{p}}\Omega_{\mathrm{p}}^{\prime}\function{v_{\mathrm{ph}}}{\omega}\function{v_{\mathrm{ph}}}{\omega^{\prime}}\function{v_{\mathrm{ph}}}{\Omega_{\mathrm{p}}}\function{v_{\mathrm{ph}}}{\Omega_{\mathrm{p}}^{\prime}}}}\op{a}^\dagger_{\omega}\op{a}^\dagger_{\omega^\prime}\op{a}_{\Omega_{\scriptsize\mathrm{p}}}\op{a}_{\Omega^\prime_{\scriptsize\mathrm{p}}}\right.\times\\
      & \times\left.\int\limits_{0}^{x}\e^{\imu\left[-\function{k}{\omega}-\function{k}{\omega^\prime}+\function{k}{\Omega_{\mathrm{p}}}+\function{k}{\Omega^\prime_{\mathrm{p}}}\right]z}+\mathrm{H.c.}\right]\diff{z}\diff{\omega}\diff{\omega^{\prime}}\diff{\Omega_{\mathrm{p}}}\diff{\Omega^{\prime}_{\mathrm{p}}}\, ,
    \end{split}
    \label{eqn-nonlinear-hamiltonian-continuum}
  \end{equation}
\end{widetext}
where we have dropped the non-resonant and fast rotating terms. The first
line in equation~\eqref{eqn-nonlinear-hamiltonian-continuum} describes the
propagation of the free photon fields. The second and third lines describe
self-phase modulation and the following two lines represent cross-phase
modulation of the strong pump tone and the weak signal spectrum. Finally, the
last two lines in equation~\eqref{eqn-nonlinear-hamiltonian-continuum}
describe the actual parametric amplification mechanism due to a nonlinear
four-wave-mixing process. The amplification of a signal with frequency
$\omega$ relies on the annihilation of two pump photons at frequencies
$\Omega_{\mathrm{p}}$ and $\Omega_{\mathrm{p}}^{\prime}$, while creating an
additional signal photon at frequency $\omega$ and an idler photon at
$\omega^{\prime}$.

In order to add losses and noise due to the imperfect substrate isolation to
our model, we introduce a phenomenological heat-bath, representing an
environmental photon field~\cite{sargent_laser_1974}. The heat-bath models
the dielectric and resistive substrate losses, which are represented by the
orange resistor in~\figref{fig-unit-cell}. The bath coupling, however, is
bidirectional such that thermally excited photons are introduced within the
system in terms of the fluctuation-dissipation
theorem~\cite{kubo_fluctuation-dissipation_1966}. The coupling constants of
the heat-bath can be related to resistive and dielectric substrate losses by
means of expanding the substrate conductivity in a Foster
representation~\cite{russer_electromagnetics_2006}, i.e., an infinite series
of harmonic oscillators~\cite{vool_introduction_2017}. These harmonic
oscillators are described by bath creation and annihilation operators
$\op{b}_m^\dagger$ and $\op{b}_m$ with closely spaced frequencies $\omega_m$.
The Hamiltonian describing the bath photon field is given by
\begin{equation}
  \op{H}_{\mathrm{bath}}=\sum_{m}\hbar\omega_m\int\limits_{0}^{\infty}\function{\op{b}_m^\dagger}{\omega}\function{\op{b}_m}{\omega}\diff{\omega}\, ,
\end{equation}
where we performed a Fourier transform of each bath mode and omitted the
zero-point energy. The exchange of energy between the heat-bath and the
system is formulated in terms of a linear coupling Hamiltonian
\begin{equation}
  \op{H}_{\mathrm{coupling}}=\sum_{m}\hbar\int\limits_{0}^{\infty}\left[\function{\kappa_m}{\omega}\function{\op{b}_m^\dagger}{\omega}\op{a}_\omega+\mathrm{H.c.}\right]\diff{\omega}\label{eqn-bath-coupling-hamiltonian}
\end{equation}
with the coupling coefficients $\function{\kappa_m}{\omega}$, describing the
interaction strength of the $m$-th bath mode and the system, respectively.
The total Hamiltonian for a dissipative JTWPA is then given by
\begin{equation}
  \op{H}_{\mathrm{total}}=\op{H}_{\mathrm{JTWPA}}+\op{H}_{\mathrm{bath}}+\op{H}_{\mathrm{coupling}}\, .\label{eqn-total-hamiltonian}
\end{equation}
Note that within this model, the heat-bath is only coupled to the weak signal
photon field $\op{a}_{\omega}$ and not to the strong pump tone
$\op{a}_{\Omega_{\scriptsize\mathrm{p}}}$. Hence, losses and noise in the
pump tone are being neglected.

The total Hamiltonian $\op{H}_{\mathrm{total}}=\op{H}_0+\op{H}_1$ is split
into an unperturbed part
\begin{equation}
  \op{H}_0=\int\limits_{0}^{\infty}\hbar\Omega_{\mathrm{p}}\op{a}^\dagger_{\Omega_{\scriptsize\mathrm{p}}}\op{a}_{\Omega_{\scriptsize\mathrm{p}}}\diff{\Omega_{\mathrm{p}}}+\int\limits_{0}^{\infty}\hbar\omega\op{a}^\dagger_{\omega}\op{a}_{\omega}\diff{\omega}\, ,
\end{equation}
which represents the free photon field propagation of the pump and weak
signal modes, and a perturbative part $\op{H}_1$ consisting of the remaining
terms. The perturbative part hence contains the descriptions for self- and
cross-phase-modulation of the strong pump field, the nonlinear
four-wave-mixing interaction, as well as losses and noise due to the
heat-bath. We use the unperturbed Hamiltonian $\op{H}_0$ to transform the
weak photon field operators to a co-rotating Heisenberg-interaction
frame~\cite{barnett_methods_1997}, given by
\begin{equation}
  \op{\tilde{a}}_\omega=\e^{-\imu\op{H}_0t/\hbar}\op{a}_\omega\e^{\imu\op{H}_0t/\hbar}=\op{a}_\omega\e^{\imu\omega t}\, .\label{eqn-co-rotating-frame}
\end{equation}
Within this co-rotating frame~\eqref{eqn-co-rotating-frame}, the total
Hamiltonian is given in terms of the perturbation Hamiltonian $\op{H}_1$
according to
\begin{equation}
  \op{\tilde{H}}_{1}=\e^{\imu\op{H}_0t/\hbar}\op{H}_1\e^{-\imu\op{H}_0t/\hbar}\, .\label{eqn-hamiltonian-interaction-picture}
\end{equation}
Equation~\eqref{eqn-hamiltonian-interaction-picture} describes the dynamics
of a JTWPA including environmental coupling in a co-rotating frame according
to the perturbations in $\op{H}_1$. However, it is difficult to derive the
temporal equations of motion of the photon field modes directly from this
Hamiltonian, due to the dispersion along the transmission line. Since the
amplification mechanism is based on parametric conversion of two pump photons
at frequencies $\Omega_{\mathrm{p}}$ and $\Omega_{\mathrm{p}}^\prime$ into a
signal photon at the desired frequency $\omega$ and an additional idler
photon with frequency $\omega^\prime$, such that
$\Omega_\mathrm{p}+\Omega_\mathrm{p}^\prime=\omega+\omega^\prime$, we assume
a strong discrete degenerate pump mode with a single pump frequency
$\Omega_\mathrm{p}=\Omega_\mathrm{p}^\prime=\omega_{\mathrm{p}}$, similar
to~\cite{grimsmo_squeezing_2017,reep_mesoscopic_2019}. Hence, we take the
monochromatic limit of the pump mode according
to~\eqref{eqn-continuous-discrete-translation}, which yields
\begin{equation}
  \sqrt{\frac{\hbar\Omega_{\mathrm{p}}}{2C^{\prime}\function{v_{\mathrm{ph}}}{\Omega_{\mathrm{p}}}}}\op{\tilde{a}}_{\Omega_{\mathrm{p}}}\to\sqrt{2\pi}\sqrt{\frac{\hbar\omega_{\mathrm{p}}}{2C^{\prime}l_{\mathrm{q}}}}\op{\tilde{a}}_{\mathrm{p}}\function{\delta}{\Omega_{\mathrm{p}}-\omega_{\mathrm{p}}}\, ,\label{eqn-discrete-pump}
\end{equation}
with the quantization length $l_\mathrm{q}$~\cite{reep_mesoscopic_2019}.
Furthermore, we adopt the asymptotic scattering limit treatment
from~\cite{grimsmo_squeezing_2017} and introduce a unitary operator $\op{U}$
to describe the time evolution of the system dynamics in a first-order
perturbation theory neglecting
time-ordering~\cite{quesada_effects_2014,quesada_time-ordering_2015},
\begin{equation}
  \function{\op{U}}{t_0,t}=\exp\left[-\frac{\imu}{\hbar}\int\limits^t_{t_{\scriptsize 0}}\function{\op{\tilde{H}}_{1}}{\tau}\diff{\tau}\right]\, ,
\end{equation}
which is in the scattering limit equal to
\begin{equation}
  \function{\op{U}}{t_0,t}\rightarrow\function{\op{U}}{-\infty,\infty}=\e^{-\imu\op{\tilde{K}}_{\mathrm{1}}/\hbar}\, .\label{eqn-time-evolution-operator}
\end{equation}
The asymptotic scattering limit implies that we are considering a JTWPA
section from $z=0$ to $z=x$ that is embedded within two semi-infinite
perfectly matched ideal linear transmission lines from $z=-\infty$ to $z=0$
and from $z=x$ to $z=\infty$~\cite{grimsmo_squeezing_2017}. The nonlinear
interaction as well as dispersion only occur in the "active" region for
$z\in\left[0,x\right]$. For the substrate loss and noise terms, we set the
coupling strength, represented by the coupling coefficients
$\function{\kappa}{\omega}$, to zero outside the JTWPA section for
$z\in\left[0,x\right]$ and neglect reflections at the interfaces. A
discussion of the scattering limit and more details about the asymptotic
approach can be found in~\cite{liscidini_asymptotic_2012}. In the asymptotic
scattering limit, together with the discrete pump
mode~\eqref{eqn-discrete-pump}, the infinite time integral
$\op{\tilde{K}}_{1}$ of the total Hamiltonian
$\op{\tilde{H}}_{1}$~\eqref{eqn-hamiltonian-interaction-picture} can be
simplified to
\begin{widetext}
  \begin{equation}
    \begin{split}
      \op{\tilde{K}}_{1}= & -\!\frac{\hbar^2\Delta{z}^3k_{\mathrm{p}}^4}{16{C^{\prime}}^2L_{\mathrm{J},0}^3I_{\mathrm{c}}^2\omega_{\mathrm{p}}^2l_{\mathrm{q}}\function{v_{\mathrm{ph}}}{\omega_{\mathrm{p}}}}\op{\tilde{a}}^\dagger_{\mathrm{p}}\op{\tilde{a}}_{\mathrm{p}}\op{\tilde{a}}^\dagger_{\mathrm{p}}\op{\tilde{a}}_{\mathrm{p}}\!\int\limits_{0}^{x}\e^{\imu\Delta{k_{\mathrm{SPM}}}z}\diff{z}\!-\!\frac{\hbar^2\Delta{z}^3k_{\mathrm{p}}^2}{4{C^{\prime}}^2L_{\mathrm{J},0}^3I_{\mathrm{c}}^2\omega_{\mathrm{p}}l_{\mathrm{q}}}\int\limits_{0}^{\infty}\frac{\function{k^2}{\omega}}{\omega\function{v_{\mathrm{ph}}}{\omega}}\op{\tilde{a}}^\dagger_{\omega}\op{\tilde{a}}_{\omega}\op{\tilde{a}}^\dagger_{\mathrm{p}}\op{\tilde{a}}_{\mathrm{p}}\!\int\limits_{0}^{x}\e^{\imu\Delta{k_{\mathrm{XPM}}}z}\diff{z}\diff{\omega}\\
      & -\!\frac{\hbar^2\Delta{z}^3k_{\mathrm{p}}^2}{8{C^{\prime}}^2L_{\mathrm{J},0}^3I_{\mathrm{c}}^2\omega_{\mathrm{p}}l_{\mathrm{q}}}\left[\int\limits_{0}^{\infty}\frac{\function{k}{\omega}\function{k}{2\omega_{\mathrm{p}}-\omega}}{\sqrt{\omega\left(2\omega_{\mathrm{p}}-\omega\right)\function{v_{\mathrm{ph}}}{\omega}\function{v_{\mathrm{ph}}}{2\omega_{\mathrm{p}}-\omega}}}\op{\tilde{a}}^\dagger_{\omega}\op{\tilde{a}}^\dagger_{2\omega_{\mathrm{p}}-\omega}\op{\tilde{a}}_{\mathrm{p}}\op{\tilde{a}}_{\mathrm{p}}\int\limits_{0}^{x}\e^{\imu\Delta{k_{\mathrm{4WM}}}z}\diff{z}\diff{\omega}+\mathrm{H.c.}\right]\\
      & +\!\int\limits_{-\infty}^{\infty}\int\limits_{0}^{\infty}\sum_{m}\hbar\omega_m\function{\op{b}_m^\dagger}{\omega}\function{\op{b}_m}{\omega}\diff{\omega}\diff{t}+\int\limits_{-\infty}^{\infty}\int\limits_{0}^{\infty}\sum_{m}\hbar\left[\function{\kappa_m}{\omega}\function{\op{b}_m^\dagger}{\omega}\op{\tilde{a}}_\omega\e^{-\imu\omega t}+\mathrm{H.c.}\right]\diff{\omega}\diff{t}\, .
    \end{split}
    \label{eqn-time-integral-interaction-hamiltonian}
  \end{equation}
\end{widetext}
The spatial phase differences $\Delta k_{\mathrm{SPM}}$, $\Delta
  k_{\mathrm{XPM}}$ and $\Delta k_{\mathrm{4WM}}$
in~\eqref{eqn-time-integral-interaction-hamiltonian} are given by
\begin{align*}
  \Delta{k_{\mathrm{SPM}}} & =k_{\mathrm{p}}-k_{\mathrm{p}}+k_{\mathrm{p}}-k_{\mathrm{p}}=0\, ,                  \\
  \Delta{k_{\mathrm{XPM}}} & =\function{k}{\omega}-\function{k}{\omega}+k_{\mathrm{p}}-k_{\mathrm{p}}=0\, ,      \\
  \Delta{k_{\mathrm{4WM}}} & =2k_{\mathrm{p}}-\function{k}{\omega}-\function{k}{2\omega_{\mathrm{p}}-\omega}\, .
\end{align*}
We now exploit the space-time symmetry of the traveling-wave amplitudes as
discussed in Section~\ref{sec-time-frame}. Hence, we introduce a reference
timeframe $t_{\mathrm{r}}$, which is related to the total JTWPA length $x$ by
$x=v_{\mathrm{r}}\,t_{\mathrm{r}}$, with a reference velocity
$v_{\mathrm{r}}$, chosen as
$v_{\mathrm{r}}=\sqrt{{\Delta{z}}/{L_{\mathrm{J},0}C^\prime}}$. The reference
time $t_\mathrm{r}$ corresponds to the propagation time of the photon field
along a transmission line of length $x$ in the dispersionless case. The
dispersion factor $\sqrt{\function{\Lambda}{\omega}}$, according
to~\eqref{eqn-dispersion-relation}, translates this reference propagation
time $t_\mathrm{r}$ to the frequency dependent total interaction time
$\function{t}{\omega}$ for each individual mode. Using the dispersion
relation~\eqref{eqn-dispersion-relation}, we can also translate between the
linear spatial phase-mismatch $\Delta{k_{\mathrm{4WM}}}$ and the linear
temporal phase-mismatch $\Delta{\Omega_{\mathrm{L}}}$ by
\begin{align}
  \Delta{k_{\mathrm{4WM}}}x & =\left[2k_{\mathrm{p}}-\function{k}{\omega}-\function{k}{2\omega_{\mathrm{p}}-\omega}\right]x\nonumber                                                                                                \\
                            & =\left[2\omega_{\mathrm{p}}\sqrt{\function{\Lambda}{\omega_{\mathrm{p}}}}-\omega\sqrt{\function{\Lambda}{\omega}}\right.\nonumber                                                                     \\
                            & \left.-\left(2\omega_{\mathrm{p}}-\omega\right)\sqrt{\function{\Lambda}{2\omega_{\mathrm{p}}-\omega}}\right]t_{\mathrm{r}}=\Delta{\Omega_\mathrm{L}}t_\mathrm{r}\, ,\label{eqn-phase-mismatch-switch}
\end{align}
where we define the frequency dependent linear phase-mismatch
$\Delta\Omega_{\mathrm{L}}$ in~\eqref{eqn-phase-mismatch-switch} by
\begin{equation*}
  \Delta\Omega_{\mathrm{L}}=2\omega_{\mathrm{p}}\sqrt{\function{\Lambda}{\omega_{\mathrm{p}}}}-\omega\sqrt{\function{\Lambda}{\omega}}-\left(2\omega_{\mathrm{p}}\!-\!\omega\right)\sqrt{\function{\Lambda}{2\omega_{\mathrm{p}}\!-\!\omega}}\, .
\end{equation*}

\section{\label{sec-heisenberg}Temporal equations of motion}

With the unitary time-evolution operator
from~\eqref{eqn-time-evolution-operator}, it holds that
$\op{\tilde{a}}_\omega=\op{U}^\dagger\op{\tilde{a}}_{\omega,0}\op{U}$, where
$\op{\tilde{a}}_{\omega,0}$ is the weak photon field annihilation operator at
some initial time $t_0$. Hence, the time-evolution of the photon field
annihilation operator $\op{\tilde{a}}_{\omega}$ in the co-rotating frame is
given by
\begin{equation}
  \begin{split}
    \pdd{\op{\tilde{a}}_\omega}{t_{\mathrm{r}}}&=\dd{\op{U}^\dagger}{t_\mathrm{r}}\op{\tilde{a}}_{\omega,0}\op{U}+\op{U}^\dagger\op{\tilde{a}}_{\omega,0}\dd{\op{U}}{t_\mathrm{r}}\\
    &=\frac{\imu}{\hbar}\dd{\op{\tilde{K}}_{1}}{t_{\mathrm{r}}}\op{U}^\dagger\op{\tilde{a}}_{\omega,0}\op{U}-\frac{\imu}{\hbar}\op{U}^\dagger\op{\tilde{a}}_{\omega,0}\op{U}\dd{\op{\tilde{K}}_{1}}{t_{\mathrm{r}}}\\
    &=\frac{\imu}{\hbar}\dd{\op{\tilde{K}}_{1}}{t_{\mathrm{r}}}\op{\tilde{a}}_{\omega}-\frac{\imu}{\hbar}\op{\tilde{a}}_{\omega}\dd{\op{\tilde{K}}_{1}}{t_{\mathrm{r}}}\\
    &=\frac{\imu}{\hbar}\commutator*{\dd{\op{\tilde{K}}_{1}}{t_{\mathrm{r}}}}{\op{\tilde{a}}_{\omega}}\, ,
  \end{split}
  \label{eqn-eom-formula}
\end{equation}
where $t_\mathrm{r}$ highlights that the time-evolution is calculated within
the reference timeframe. Thus, within the reference timeframe $t_\mathrm{r}$,
the system Hamiltonian in the Heisenberg picture is given in terms of the
derivative of $\op{\tilde{K}}_{1}$ with respect to $t_\mathrm{r}$, where the
length $x$ of the JTWPA has been replaced by $x=v_\mathrm{r}\,t_\mathrm{r}$,
i.e.\
\begin{widetext}
  \begin{equation}
    \begin{split}
      \dd{\op{\tilde{K}}_{1}}{t_{\mathrm{r}}}= & -\frac{\hbar^2\Delta{z}^3k_{\mathrm{p}}^4\sqrt{\function{\Lambda}{\omega_{\mathrm{p}}}}}{16{C^{\prime}}^2L_{\mathrm{J},0}^3I_{\mathrm{c}}^2\omega_{\mathrm{p}}^2l_{\mathrm{q}}}\op{\tilde{a}}^\dagger_{\mathrm{p}}\op{\tilde{a}}_{\mathrm{p}}\op{\tilde{a}}^\dagger_{\mathrm{p}}\op{\tilde{a}}_{\mathrm{p}}-\frac{\hbar^2\Delta{z}^3k_{\mathrm{p}}^2}{4{C^{\prime}}^2L_{\mathrm{J},0}^3I_{\mathrm{c}}^2\omega_{\mathrm{p}}l_{\mathrm{q}}}\int\limits_{0}^{\infty}\frac{\function{k^2}{\omega}\sqrt{\function{\Lambda}{\omega}}}{\omega}\op{\tilde{a}}^\dagger_{\omega}\op{\tilde{a}}_{\omega}\op{\tilde{a}}^\dagger_{\mathrm{p}}\op{\tilde{a}}_{\mathrm{p}}\diff{\omega}\\
      & -\frac{\hbar^2\Delta{z}^3k_{\mathrm{p}}^2}{8{C^{\prime}}^2L_{\mathrm{J},0}^3I_{\mathrm{c}}^2\omega_{\mathrm{p}}l_{\mathrm{q}}}\left[\int\limits_{0}^{\infty}\frac{\function{k}{\omega}\function{k}{2\omega_{\mathrm{p}}-\omega}\left[\function{\Lambda}{\omega}\function{\Lambda}{2\omega_{\mathrm{p}}-\omega}\right]^{\frac{1}{4}}}{\sqrt{\omega\left(2\omega_{\mathrm{p}}-\omega\right)}}\op{\tilde{a}}^\dagger_{\omega}\op{\tilde{a}}^\dagger_{2\omega_{\mathrm{p}}-\omega}\op{\tilde{a}}_{\mathrm{p}}\op{\tilde{a}}_{\mathrm{p}}\,\e^{\imu\Delta{\Omega_{\mathrm{L}}}t_{\mathrm{r}}}\diff{\omega}+\mathrm{H.c.}\right]\\
      & +\sum_{m}\hbar\omega_m\sqrt{\function{\Lambda}{\omega_m}}\int\limits_{0}^{\infty}\function{\op{b}_m^\dagger}{\omega}\function{\op{b}_m}{\omega}\diff{\omega}+\sum_{m}\hbar\sqrt{\function{\Lambda}{\omega_m}}\int\limits_{0}^{\infty}\left[\function{\kappa_m}{\omega}\function{\op{b}_m^\dagger}{\omega}\op{\tilde{a}}_\omega\e^{-\imu\omega\sqrt{\function{\Lambda}{\omega_m}}t_{\mathrm{r}}}+\mathrm{H.c.}\right]\diff{\omega}\, .
    \end{split}
    \label{eqn-hamiltonian-diff-time}
  \end{equation}
\end{widetext}

Within the reference timeframe $t_\mathrm{r}$, equations of motion for the
system and bath operators can be obtained from~\eqref{eqn-eom-formula}
and~\eqref{eqn-hamiltonian-diff-time}, according to
\begin{align}
  \pdd{\op{\tilde{a}}_{\mathrm{p}}}{t_{\mathrm{r}}} & =\,\frac{\imu\hbar\Delta
  z^3k^4_{\mathrm{p}}\sqrt{\function{\Lambda}{\omega_{\mathrm{p}}}}}{8{C^{\prime}}^2L^3_{\mathrm{J},0}I^2_{\mathrm{c}}\omega_{\mathrm{p}}^2l_{\mathrm{q}}}\op{\tilde{a}}^\dagger_{\mathrm{p}}\op{\tilde{a}}_{\mathrm{p}}\op{\tilde{a}}_{\mathrm{p}}\, ,\label{eqn-heom-first}                                                                                                                                                   \\
  \pdd{\op{\tilde{a}}_{\omega}}{t_{\mathrm{r}}}     & =\,\frac{\imu\hbar\Delta
  z^3k^2_{\mathrm{p}}\function{k^2}{\omega}\sqrt{\function{\Lambda}{\omega}}}{4{C^{\prime}}^2L^3_{\mathrm{J},0}I^2_{\mathrm{c}}\omega_{\mathrm{p}}\omega l_{\mathrm{q}}}\op{\tilde{a}}^\dagger_{\mathrm{p}}\op{\tilde{a}}_{\mathrm{p}}\op{\tilde{a}}_{\omega}\nonumber                                                                                                                                                          \\
                                                    & \quad+\frac{\imu\hbar\Delta z^3k^2_{\mathrm{p}}\function{k}{\omega}\function{k}{2\omega_{\mathrm{p}}-\omega}\left[\function{\Lambda}{\omega}\function{\Lambda}{2\omega_{\mathrm{p}}-\omega}\right]^{\frac{1}{4}}}{8{C^{\prime}}^2L^3_{\mathrm{J},0}I^2_{\mathrm{c}}\omega_{\mathrm{p}}\sqrt{\omega\left(2\omega_{\mathrm{p}}-\omega\right)}l_{\mathrm{q}}}\times\nonumber \\
                                                    & \quad\times\op{\tilde{a}}_{\mathrm{p}}\op{\tilde{a}}_{\mathrm{p}}\op{\tilde{a}}^\dagger_{2\omega_{\mathrm{p}}-\omega}\,\e^{\imu\Delta \Omega_{\mathrm{L}}t_{\mathrm{r}}}\nonumber                                                                                                                                                                                         \\
                                                    & \quad-\imu\sum_{m}\function{\kappa_m}{\omega}\sqrt{\function{\Lambda}{\omega_m}}\,\op{b}_m\e^{\imu\omega\sqrt{\function{\Lambda}{\omega_m}}t_{\mathrm{r}}}\, ,                                                                                                                                                                                                            \\
  \pdd{\op{b}_m}{t_{\mathrm{r}}}                    & =\,-\imu\omega_m\sqrt{\function{\Lambda}{\omega_m}}\,\op{b}_m\nonumber                                                                                                                                                                                                                                                                                                    \\
                                                    & \quad-\imu\function{\kappa_m}{\omega}\sqrt{\function{\Lambda}{\omega_m}}\,\op{\tilde{a}}_{\omega}\e^{-\imu\omega\sqrt{\function{\Lambda}{\omega_m}}t_{\mathrm{r}}}\, .\label{eqn-heom-last}
\end{align}
As the pump mode is by orders of magnitude stronger than the weak signal
photon field, we furthermore assume that the pump tone can be approximated by
a classical mode amplitude, i.e.\
\begin{equation}
  \sqrt{\frac{\hbar}{2C^{\prime}\omega_{\mathrm{p}}l_{\mathrm{q}}}}\op{\tilde{a}}_{\mathrm{p}}\to-\frac{\imu}{2}A_{\mathrm{p}}\, .\label{eqn-classical-pump-mode}
\end{equation}
Consequently, we also neglect pump depletion, i.e.\ photons that are being
annihilated in the pump mode to create a signal and an idler photon by
four-wave-mixing will not deplete the pump mode~\cite{reep_mesoscopic_2019}.
This is a bold assumption, as it violates energy conservation, but it can be
justified by the huge difference in the number of photons in the weak signal
field compared to the strong classical pump tone. However, it allows us to
construct analytic solutions for the temporal evolution of the photon field,
as we will see in the following. We also neglect leakage of pump power to
higher harmonics~\cite{zorin_josephson_2016}. Within our model, terms of
smaller than second order in the pump amplitude are neglected.

The classical pump approximation~\eqref{eqn-classical-pump-mode}, while
neglecting pump depletion~\cite{reep_mesoscopic_2019}, drastically simplifies
the resulting equation of motion. Equations~\eqref{eqn-heom-first}
to~\eqref{eqn-heom-last}, together with~\eqref{eqn-classical-pump-mode}
result in a set of approximated first-order temporal Heisenberg equations,
given by
\begin{align}
  \pdd{A_{\mathrm{p}}}{t_{\mathrm{r}}}          & =\,\imu\theta_{\mathrm{p}}\sqrt{\function{\Lambda}{\omega_{\mathrm{p}}}}A_{\mathrm{p}}\, ,
  \label{eqn-heom-pump}                                                                                                                                                                                                                                                                                                 \\
  \pdd{\op{\tilde{a}}_{\omega}}{t_{\mathrm{r}}} & =\,\imu\function{\theta}{\omega}\sqrt{\function{\Lambda}{\omega}}\,\op{\tilde{a}}_{\omega}\nonumber                                                                                                                                                                   \\
                                                & \quad-\imu\chi^{\prime}A_{\mathrm{p}}^2\left[\function{\Lambda}{\omega}\function{\Lambda}{2\omega_{\mathrm{p}}-\omega}\right]^{\frac{1}{4}}\,\op{\tilde{a}}^\dagger_{2\omega_{\scriptsize\mathrm{p}}-\omega}\e^{\imu\Delta\Omega_{\mathrm{L}}t_{\mathrm{r}}}\nonumber \\
                                                & \quad-\imu\sum_{m}\function{\kappa_m}{\omega}\sqrt{\function{\Lambda}{\omega_m}}\,\op{b}_m\e^{\imu\omega\sqrt{\function{\Lambda}{\omega_m}}t_{\mathrm{r}}}\, ,
  \label{eqn-heom-signal}                                                                                                                                                                                                                                                                                               \\
  \pdd{\op{b}_m}{t_\mathrm{r}}                  & =\,-\imu\omega_m\sqrt{\function{\Lambda}{\omega_m}}\,\op{b}_m\nonumber                                                                                                                                                                                                \\
                                                & \quad-\imu\function{\kappa_m}{\omega}\sqrt{\function{\Lambda}{\omega_m}}\op{\tilde{a}}_{\omega}\e^{-\imu\omega\sqrt{\function{\Lambda}{\omega_m}}t_{\mathrm{r}}}\, .
  \label{eqn-heom-bath}
\end{align}
Self-phase-modulation of the strong monochromatic classical pump amplitude is
described by means of $\theta_{\mathrm{p}}$ and cross-phase-modulation is
included in the weak signal modes through $\function{\theta}{\omega}$. The
coefficient for cross-phase-modulation $\function{\theta}{\omega}$ is given
by
\begin{equation}
  \function{\theta}{\omega}=\,\frac{k_{\mathrm{p}}^2\Delta{z}^2\function{\Lambda}{\omega}\omega}{8L_{\mathrm{J},0}^2I_{\mathrm{c}}^2}\abs{A_{\mathrm{p,0}}}^2\, ,
\end{equation}
while the one for self-phase modulation
$\theta_{\mathrm{p}}=\function{\theta}{\omega_{\mathrm{p}}}/2$. Here,
$\abs{A_{\mathrm{p,0}}}$ is the initial magnitude of the monochromatic pump
tone at the amplifier input. The four-wave-mixing interaction strength
$\chi^{\prime}$ in \eqref{eqn-heom-signal} was summarized as
\begin{equation}
  \chi^{\prime}=\frac{k_{\mathrm{p}}^2\Delta{z}^2\sqrt{\function{\Lambda}{\omega}\function{\Lambda}{2\omega_{\mathrm{p}}-\omega}}\sqrt{\omega\left(2\omega_{\mathrm{p}}-\omega\right)}}{16L_{\mathrm{J},0}^2I_{\mathrm{c}}^2}\, .
\end{equation}
The wave number $k_\mathrm{p}$ can be obtained from the reference velocity
$v_{\mathrm{r}}$, considering the dispersion factor given
in~\eqref{eqn-dispersion-relation}
\begin{equation}
  \function{k}{\omega}=\omega\sqrt{\frac{L_{\mathrm{J},0}}{\Delta{z}}C'\function{\Lambda}{\omega}}\, ,
\end{equation}
where $k_\mathrm{p}=\function{k}{\omega_{\mathrm{p}}}$. A detailed derivation
of the dispersion relation and the wave number of a dissipative transmission
line is given in~\cite{zurita_lossy_2006}.

As we neglect pump depletion and losses, \eqref{eqn-heom-pump} is decoupled
from the rest of the system and can be solved independently. Hence, by
formally integrating~\eqref{eqn-heom-pump}, the temporal evolution of the
classical pump amplitude is given by
\begin{equation}
  A_{\mathrm{p}}=A_{\mathrm{p},0}\,\e^{\imu\theta_{\mathrm{p}}\sqrt{\function{\Lambda}{\omega_{\mathrm{p}}}}t_{\mathrm{r}}}\, .\label{eqn-solution-pump}
\end{equation}
Therefore, the amplitude of the monochromatic strong undepleted classical
pump mode in~\eqref{eqn-solution-pump} can be considered constant over space
and time, $\abs{A_{\mathrm{p}}}=\abs{A_{\mathrm{p},0}}$.

The bath mode operators $\op{b}_m$ can be expresses in terms of the initial
bath amplitudes $\op{b}_{m,0}$ and the mode operators
$\op{\tilde{a}}_{\omega}$ by formally
integrating~\eqref{eqn-heom-bath}~\cite{jirauschek_nonlinear_2012}
\begin{equation}
  \begin{split}
    \op{b}_m & =\op{b}_{m,0}\e^{-\imu\omega_m\sqrt{\function{\Lambda}{\omega_m}}t_\mathrm{r}}-\imu\function{\kappa_m}{\omega}\sqrt{\function{\Lambda}{\omega_m}}\times \\
    &\quad\times\int\limits_{0}^{t_{\mathrm{r}}}\function{\op{\tilde{a}}_{\omega}}{t_{\mathrm{r}}-\tau}\e^{-\imu\omega\sqrt{\function{\Lambda}{\omega_m}}\left(t_{\mathrm{r}}-\tau\right)}\e^{-\imu\omega_m\sqrt{\function{\Lambda}{\omega_m}}\tau}\diff{\tau}\, .
  \end{split}
  \label{eqn-integral-bath}
\end{equation}
Inserting~\eqref{eqn-integral-bath} and the solution for the classical pump
amplitude~\eqref{eqn-solution-pump} into the equation of motion for the weak
photon field operators $\op{\tilde{a}}_{\omega}$, we obtain
\begin{widetext}
  \begin{equation}
    \begin{split}
      \pdd{\op{\tilde{a}}_{\omega}}{t_{\mathrm{r}}}&=\imu\function{\theta}{\omega}\sqrt{\function{\Lambda}{\omega}}\,\op{\tilde{a}}_{\omega}-\imu\chi^{\prime}A_{\mathrm{p},0}^2\left[\function{\Lambda}{\omega}\function{\Lambda}{2\omega_{\mathrm{p}}-\omega}\right]^{\frac{1}{4}}\op{\tilde{a}}_{2\omega_{\mathrm{p}}-\omega}^{\dagger}\e^{\imu\left[2\theta_{\mathrm{p}}\sqrt{\function{\Lambda}{\omega_{\mathrm{p}}}}+\Delta\Omega_{\mathrm{L}}\right]t_{\mathrm{r}}}\\
      &\quad-\imu\sum_{m}\function{\kappa_m}{\omega}\sqrt{\function{\Lambda}{\omega_m}}\,\op{b}_{m,0}\,\e^{-\imu\left[\omega_m\sqrt{\function{\Lambda}{\omega_m}}-\omega\sqrt{\function{\Lambda}{\omega_m}}\right]t_{\mathrm{r}}}\\
      &\quad-\sum_{m}\function{\kappa_m^2}{\omega}\function{\Lambda}{\omega_m}\int\limits_{0}^{t_{\mathrm{r}}}\function{\op{\tilde{a}}_\omega}{t_{\mathrm{r}}-\tau}\e^{-\imu\left[\omega_m\sqrt{\function{\Lambda}{\omega_m}}-\omega\sqrt{\function{\Lambda}{\omega_m}}\right]\tau}\diff{\tau}\, .
    \end{split}
    \label{eqn-heom-rotating-frame}
  \end{equation}
\end{widetext}
Here, the first line in~\eqref{eqn-heom-rotating-frame} expresses the
evolution of the unperturbed system, together with the four-wave-mixing
interaction due to the Josephson nonlinearity. The second and the third line
represent fluctuation and dissipation, respectively.

In a next step, we assume a memory-less Markovian system, i.e.\ the
interference time is much smaller than the time over which significant
changes in phase and amplitude of $\op{a}_{\omega}$
occur~\cite{sargent_laser_1974,weisskopf_linienbreite_1930}. In the context
of Josephson parametric amplifiers, a similar derivation has already been
given in~\cite{kaiser_quantum_2017}, however, there it was done for discrete
mode operators in a lumped-element JPA resonator. Distributed losses in
quantum traveling-wave parametric amplifiers have been studied
in~\cite{houde_loss_2019} and~\cite{peng_floquet-mode_2022}, using a
continuous input-output theory. As the bath mode frequencies $\omega_m$ are
closely spaced, we replace the sum over $m$ bath modes by an integral over a
continuous frequency argument $\Omega_{\mathrm{b}}$
\begin{equation}
  \sum\limits_{m}\rightarrow\int\limits_{-\infty}^{\infty}\diff{\Omega_{\mathrm{b}}}\function{\mathcal{D}}{\Omega_{\mathrm{b}}}\, ,
\end{equation}
where $\function{\mathcal{D}}{\Omega_{\mathrm{b}}}$ is the one-dimensional
density of the states. Here, the heat-bath is modeled as a distributed
resistance in terms of a closely spaced ensemble of harmonic oscillators,
that is coupled to the system in each unit cell. We define
frequency-dependent damping factors by
\begin{equation}
  \function{\gamma}{\omega}=\,2\pi\function{\mathcal{D}}{\omega}\function{\kappa^2}{\omega}\sqrt{\function{\Lambda}{\omega}}\, ,
\end{equation}
where the translation factor $\sqrt{\function{\Lambda}{\omega}}$ ensures that
the correct propagation time/length is taken into account for each mode when
calculating the losses. In order to further simplify the equation of motion
of the weak photon field operators, we again switch to a co-rotating frame
$\op{A}_{\omega}=\op{\tilde{a}}_{\omega}\exp{\left\lbrace-\imu\left[\function{\theta}{\omega}\sqrt{\function{\Lambda}{\omega}}+\Delta{\Omega}/2\right]t_{\mathrm{r}}\right\rbrace}$,
with the frequency-dependent total phase mismatch
\begin{equation}
  \begin{split}
    \Delta{\Omega}&=2\left[\omega_{\mathrm{p}}+\theta_{\mathrm{p}}\right]\sqrt{\function{\Lambda}{\omega_{\mathrm{p}}}}-\left[\omega+\function{\theta}{\omega}\right]\sqrt{\function{\Lambda}{\omega}}\\
    &\quad-\left[2\omega_{\mathrm{p}}-\omega+\function{\theta}{2\omega_{\mathrm{p}}-\omega}\right]\sqrt{\function{\Lambda}{2\omega_{\mathrm{p}}-\omega}}\, .
  \end{split}
  \label{eqn-total-phase-mismatch}
\end{equation}
Within this co-rotating frame, we obtain a system of two coupled first-order
inhomogeneous differential equations for the weak photon field operators
$\op{A}_{\omega}$, one centered at the signal frequency $\omega$, and one at
the idler frequency $2\omega_{\mathrm{p}}-\omega$. The resulting system of
equations can be written in matrix form as
\begin{widetext}
  \begin{equation}
    \begin{bmatrix}
      \pdd{\op{A}_{\omega}}{t_\mathrm{r}} \\
      \pdd{\op{A}^\dagger_{2\omega_{\mathrm{p}}-\omega}}{t_\mathrm{r}}
    \end{bmatrix}=\begin{bmatrix}
      -\frac{\function{\gamma}{\omega}}{2}-\frac{\imu\Delta{\Omega}}{2}
       & -\imu\chi^{\prime}A_{\mathrm{p},0}^2\left[\function{\Lambda}{\omega}\function{\Lambda}{2\omega_{\mathrm{p}}-\omega}\right]^{\frac{1}{4}} \\
      \imu\chi^{\prime\ast}{A_{\mathrm{p},0}^{\ast}}^2\left[\function{\Lambda}{\omega}\function{\Lambda}{2\omega_{\mathrm{p}}-\omega}\right]^{\frac{1}{4}}
       & -\frac{\function{\gamma}{2\omega_{\mathrm{p}}-\omega}}{2}+\frac{\imu\Delta{\Omega}}{2}                                                   \\
    \end{bmatrix}\begin{bmatrix}
      \op{A}_{\omega}                              \\
      \op{A}^\dagger_{2\omega_{\mathrm{p}}-\omega} \\
    \end{bmatrix} +\begin{bmatrix}
      \function{\op{f}}{\omega}                              \\
      \function{\op{f}^\dagger}{2\omega_{\mathrm{p}}-\omega} \\
    \end{bmatrix}\, .
    \label{eqn-heom-matrix}
  \end{equation}
\end{widetext}
While losses are directly incorporated in the system matrix
in~\eqref{eqn-heom-matrix} through the damping factors
$\function{\gamma}{\omega}$, noise is included by the inhomogeneous
fluctuation operators $\function{\op{f}}{\omega}$ and
$\function{\op{f}}{2\omega_{\mathrm{p}}-\omega}$. These fluctuation operators
are given by
\begin{equation}
  \begin{split}
    \function{\op{f}}{\omega} & =-\imu\sum_{m}\function{\kappa_m}{\omega}\sqrt{\function{\Lambda}{\omega}}\function{\op{b}_{m,0}}{\omega}\times\\
    & \quad\times\e^{-\imu\left[\left(\omega_m-\omega\right)\sqrt{\function{\Lambda}{\omega_m}}+\function{\theta}{\omega}\sqrt{\function{\Lambda}{\omega}}+\frac{\Delta{\Omega}}{2}\right] t_{\mathrm{r}}}\, .
  \end{split}
  \label{eqn-fluctuation-operator}
\end{equation}

\section{\label{sec-analytic-solution}Analytic solution for the photon field operator}

The inhomogeneous system of differential equations~\eqref{eqn-heom-matrix}
can be solved analytically by standard methods. By evaluating an eigenvector
basis of the system matrix in~\eqref{eqn-heom-matrix}, the time evolution of
the weak signal photon field annihilation operator $\op{A}_{\omega}$ is found
as
\begin{widetext}
  \begin{equation}
    \begin{split}
      \op{A}_{\omega}&=\left[\cosh{\left(g t_{\mathrm{r}}\right)}-\frac{\function{\gamma}{\omega}-\function{\gamma}{2\omega_{\mathrm{p}}-\omega}+2\imu\Delta{\Omega}}{4g}\sinh{\left(gt_{\mathrm{r}}\right)}\right]\op{A}_{\omega,0}\e^{-\frac{\function{\gamma}{\omega}+\function{\gamma}{2\omega_{\mathrm{p}}-\omega}}{4}t_{\mathrm{r}}}\\
      &\quad-\frac{\imu\chi^{\prime}A^2_{\mathrm{p,0}}\left[\function{\Lambda}{\omega}\function{\Lambda}{2\omega_{\mathrm{p}}-\omega}\right]^{\frac{1}{4}}}{g}\sinh{\left(g t_{\mathrm{r}}\right)}\op{A}^\dagger_{2\omega_{\mathrm{p}}-\omega,0}\e^{-\frac{\function{\gamma}{\omega}+\function{\gamma}{2\omega_{\mathrm{p}}-\omega}}{4}t_{\mathrm{r}}}\\
      &\quad+\int\limits_{0}^{t_{\mathrm{r}}}\left\lbrace\cosh{\left[g\cdot\left(t_{\mathrm{r}}-\tau\right)\right]}-\frac{\function{\gamma}{\omega}-\function{\gamma}{2\omega_{\mathrm{p}}-\omega}+2\imu\Delta{\Omega}}{4g}\sinh{\left[g\cdot\left(t_{\mathrm{r}}-\tau\right)\right]}\right\rbrace\function{\op{f}}{\omega}\,\e^{-\frac{\function{\gamma}{\omega}+\function{\gamma}{2\omega_{\mathrm{p}}-\omega}}{4}\left(t_{\mathrm{r}}-\tau\right)}\diff{\tau} \\
      &\quad-\int\limits_{0}^{t_{\mathrm{r}}}\frac{\imu\chi^{\prime}A_{\mathrm{p,0}}^2\left[\function{\Lambda}{\omega}\function{\Lambda}{2\omega_{\mathrm{p}}-\omega}\right]^{\frac{1}{4}}}{g}\sinh{\left[g\cdot\left(t_{\mathrm{r}}-\tau\right)\right]}\function{\op{f}^{\dagger}}{2\omega_{\mathrm{p}}-\omega}\,\e^{-\frac{\function{\gamma}{\omega}+\function{\gamma}{2\omega_{\mathrm{p}}-\omega}}{4}\left(t_{\mathrm{r}}-\tau\right)}\diff{\tau}\\
      &=\function{\zeta_1}{\omega,t_\mathrm{r}}\op{A}_{\omega,0}\e^{-\frac{\function{\gamma}{\omega}+\function{\gamma}{2\omega_{\mathrm{p}}-\omega}}{4}t_{\mathrm{r}}}+\function{\zeta_2}{\omega,t_\mathrm{r}}\op{A}^\dagger_{2\omega_{\mathrm{p}}-\omega,0}\e^{-\frac{\function{\gamma}{\omega}+\function{\gamma}{2\omega_{\mathrm{p}}-\omega}}{4}t_{\mathrm{r}}}\\
      &\quad+\int\limits_{0}^{t_{\mathrm{r}}}\function{\zeta_1}{\omega,t_\mathrm{r}-\tau}\function{\op{f}}{\omega}\,\e^{-\frac{\function{\gamma}{\omega}+\function{\gamma}{2\omega_{\mathrm{p}}-\omega}}{4}\left(t_{\mathrm{r}}-\tau\right)}\diff{\tau}+\int\limits_{0}^{t_{\mathrm{r}}}\function{\zeta_2}{\omega,t_\mathrm{r}-\tau}\function{\op{f}^{\dagger}}{2\omega_{\mathrm{p}}-\omega}\,\e^{-\frac{\function{\gamma}{\omega}+\function{\gamma}{2\omega_{\mathrm{p}}-\omega}}{4}\left(t_{\mathrm{r}}-\tau\right)}\diff{\tau}\, ,
    \end{split}
    \label{eqn-solution-signal-operator}
  \end{equation}
  %
  %
  with the gain rate
  \begin{equation}
    g=\sqrt{\left[\frac{\function{\gamma}{\omega}-\function{\gamma}{2\omega_{\mathrm{p}}-\omega}+2\imu\Delta{\Omega}}{4}\right]^2+\abs{\chi^{\prime}}^2\abs{A_{\mathrm{p,0}}^2}^2\sqrt{\function{\Lambda}{\omega}\function{\Lambda}{2\omega_{\mathrm{p}}-\omega}}}\, .
    \label{eqn-gain-factor}
  \end{equation}
  The signal and idler evolution functions
  $\function{\zeta_1}{\omega,t_{\mathrm{r}}}$ and
  $\function{\zeta_2}{\omega,t_{\mathrm{r}}}$ are given by
  \begin{align}
    \function{\zeta_1}{\omega,t_{\mathrm{r}}} & =\cosh\!{\left(g t_{\mathrm{r}}\right)}-\function{\eta}{\omega}\sinh\!{\left(g t_{\mathrm{r}}\right)}\, , \\
    \function{\zeta_2}{\omega,t_{\mathrm{r}}} & =\function{\rho}{\omega}\sinh{\left(g t_{\mathrm{r}}\right)}\, ,
  \end{align}
  with
  \begin{align}
    \function{\eta}{\omega} & =\frac{\function{\gamma}{\omega}-\function{\gamma}{2\omega_{\mathrm{p}}-\omega}+2\imu\Delta{\Omega}}{4g}\, ,\label{eqn-eta}                                            \\
    \function{\rho}{\omega} & =-\frac{\imu\chi^{\prime}A^2_{\mathrm{p,0}}\left[\function{\Lambda}{\omega}\function{\Lambda}{2\omega_{\mathrm{p}}-\omega}\right]^{\frac{1}{4}}}{g}\, .\label{eqn-rho}
  \end{align}
\end{widetext}
The first line in~\eqref{eqn-solution-signal-operator} represents the
amplification of the weak photon field at frequencies around $\omega$ due to
four-wave-mixing. The contributions of the added noise due to the
down-conversion of energy from the idler field at
$2\omega_{\mathrm{p}}-\omega$ is shown in the second line. Both contributions
experience exponential damping by the damping factors
$\function{\gamma}{\omega}$ and
$\function{\gamma}{2\omega_{\mathrm{p}}-\omega}$, which are due to losses
within the dielectric substrate. The next two lines describe the evolution of
thermal fluctuations within the quantum system due to the lossy environment.
While the integrals in the last two lines
of~\eqref{eqn-solution-signal-operator} could be easily evaluated, we refrain
from doing so here for the sake of a more streamlined derivation of the
second-order moments later on. It can be seen that photons originating from
thermal fluctuations experience similar amplification and damping compared to
the weak signal photon field.

The gain factor $g$ from~\eqref{eqn-gain-factor} describes the amplification
of the system per unit time within the reference timeframe
$t=t_{\mathrm{r}}$. According to~\eqref{eqn-gain-factor}, high gain can be
achieved by increasing the four-wave-mixing interaction strength
$\chi^{\prime}$ as well as the pump amplitude $A_{\mathrm{p},0}$. Thus,
JTWPAs are typically operated with pump powers just below the limits set by
the critical current of the Josephson
junctions~\cite{zorin_flux-driven_2019}.

Let us note here, that the analytic result must represent a bosonic mode
annihilation operator. Thus, one can verify whether the analytic solution
$\op{A}_\omega$ satisifies the bosonic commutation relation. A
straightforward calculation shows that the commutation relation is indeed
satisfied for the given result, i.e.\
\begin{equation}
  \commutator{\op{A}_\omega}{\op{A}^\dagger_{\omega^\prime}}=\op{A}_\omega\op{A}^\dagger_{\omega^\prime}-\op{A}^\dagger_{\omega^\prime}\op{A}_\omega=\function{\delta}{\omega-\omega^\prime}\, .\label{eqn-commutation-relation}
\end{equation}
In the dissipationless limit, we recover the well-known condition
$\abs{\function{\zeta_1}{\omega,t_{\mathrm{r}}}}^2-\abs{\function{\zeta_2}{\omega,t_{\mathrm{r}}}}^2=1$
from \cite{grimsmo_squeezing_2017,fasolo_bimodal_2022}, which gives us
further confirmation.

\section{\label{sec-gain-evolution}Gain spectrum and temporal evolution}

For both cases with and without RPM, the expected photon number, and hence
the gain of the JTWPA, can be obtained by integrating the spectral density,
given by the auto- and cross-correlation functions of the photon field
operators
\begin{widetext}
  \begin{equation}
    \begin{split}
      \varexpect{\op{A}^\dagger_{\omega}\op{A}_{\omega^\prime}}&=\left[\function{\zeta_1^\ast}{\omega,t_{\mathrm{r}}}\function{\zeta_1}{\omega^\prime,t_{\mathrm{r}}}\varexpect{\op{A}^\dagger_{\omega,0}\op{A}_{\omega^\prime,0}} +\function{\zeta_2^\ast}{\omega,t_{\mathrm{r}}}\function{\zeta_2}{\omega^\prime,t_{\mathrm{r}}}\varexpect{\op{A}_{2\omega_{\mathrm{p}}-\omega,0}\op{A}^\dagger_{2\omega_{\mathrm{p}}-\omega^\prime,0}}\right.\\
      &\quad+\left.\function{\zeta_1^\ast}{\omega,t_{\mathrm{r}}}\function{\zeta_2}{\omega^\prime,t_{\mathrm{r}}}\varexpect{\op{A}^\dagger_{\omega,0}\op{A}^\dagger_{2\omega_{\mathrm{p}}-\omega^{\prime},0}}+\function{\zeta_2^\ast}{\omega,t_{\mathrm{r}}}\function{\zeta_1}{\omega^{\prime},t_{\mathrm{r}}}\varexpect{\op{A}_{2\omega_{\mathrm{p}}-\omega,0}\op{A}_{\omega^{\prime},0}}\right]\e^{-\frac{\function{\gamma}{\omega}+\function{\gamma}{2\omega_{\mathrm{p}}-\omega}+\function{\gamma}{\omega^\prime}+\function{\gamma}{2\omega_{\mathrm{p}}-\omega^\prime}}{4}t_{\mathrm{r}}}\\
      &\quad+\int\limits_{0}^{t_{\mathrm{r}}}\int\limits_{0}^{t_{\mathrm{r}}}\left[\function{\zeta_1^\ast}{\omega,t_{\mathrm{r}}-\tau}\function{\zeta_1}{\omega^\prime,t_{\mathrm{r}}-\tau^\prime}\varexpect{\function{\op{f}^\dagger}{\omega}\function{\op{f}}{\omega^\prime}}+\function{\zeta_2^\ast}{\omega,t_{\mathrm{r}}-\tau}\function{\zeta_2}{\omega^\prime,t_{\mathrm{r}}-\tau^\prime}\varexpect{\function{\op{f}}{2\omega_{\mathrm{p}}-\omega}\function{\op{f}^\dagger}{2\omega_{\mathrm{p}}-\omega^\prime}}\right.\\
      &\qquad\qquad\!\!+\left.\function{\zeta_1^\ast}{\omega,t_{\mathrm{r}}-\tau}\function{\zeta_2}{\omega^\prime,t_{\mathrm{r}}-\tau^\prime}\varexpect{\function{\op{f}^\dagger}{\omega}\function{\op{f}^\dagger}{2\omega_{\mathrm{p}}-\omega^\prime}}+\function{\zeta_2^\ast}{\omega,t_{\mathrm{r}}-\tau}\function{\zeta_1}{\omega^\prime,t_{\mathrm{r}}-\tau^\prime}\varexpect{\function{\op{f}}{2\omega_{\mathrm{p}}-\omega}\function{\op{f}}{\omega^\prime}}\right]\times\\
      &\qquad\qquad\!\!\times\e^{-\frac{\function{\gamma}{\omega}+\function{\gamma}{2\omega_{\mathrm{p}}-\omega}}{4}\left(t_{\mathrm{r}}-\tau\right)}\e^{-\frac{\function{\gamma}{\omega^\prime}+\function{\gamma}{2\omega_{\mathrm{p}}-\omega^\prime}}{4}\left(t_{\mathrm{r}}-\tau^\prime\right)}\diff{\tau}\diff{\tau^\prime}\, .
    \end{split}
    \label{eqn-correlation}
  \end{equation}
\end{widetext}
The correlation functions of the noise operators $\function{\op{f}}{\omega}$
in~\eqref{eqn-correlation} are given by
\begin{equation}
  \varexpect{\function{\hat{b}^\dagger_{m,0}}{\omega}\function{\hat{b}_{n,0}}{\omega^\prime}}=\function{\bar{n}}{\omega_m}\delta_{mn}\function{\delta}{\omega-\omega^\prime}\, ,\label{eqn-noise-operator-correlation}
\end{equation}
where $\function{\bar{n}}{\omega_m}$ is the expected occupation of the $m$-th
bath mode. Assuming that the heat-bath is a photon reservoir in thermal
equilibrium at temperature $T$, the expected photon number in the $m$-th bath
mode is given by Bose-Einstein statistics
\begin{equation}
  \function{\bar{n}}{\omega_m}=\frac{1}{\e^{\hbar\omega_m/\left(k_{\mathrm{B}}T\right)}-1}\, ,\label{eqn-occupation-number}
\end{equation}
where $k_{\mathrm{B}}$ is Boltzmann's constant.

We assume that the input photon field has discrete mode spectral densities,
given by
\begin{equation}
  \varexpect{\op{A}^\dagger_{\omega,0}\op{A}_{\omega^{\prime},0}}=N_{\mathrm{s},0}\,\function{\delta}{\omega-\omega^{\prime}}\, ,
\end{equation}
\begin{equation}
  \varexpect{\op{A}_{2\omega_\mathrm{p}-\omega,0}\op{A}^\dagger_{2\omega_\mathrm{p}-\omega^{\prime},0}}=\left(N_{\mathrm{i},0}+1\right)\,\function{\delta}{\omega-\omega^{\prime}}\, ,
\end{equation}
with initial correlations
\begin{equation}
  \varexpect{\op{A}_{2\omega_\mathrm{p}-\omega,0}\op{A}_{\omega^{\prime},0}}=\varexpect{\op{A}_{\omega,0}^\dagger\op{A}^\dagger_{2\omega_\mathrm{p}-\omega^{\prime},0}}^\ast=C_{\mathrm{s}\mathrm{i},0}\,\function{\delta}{\omega-\omega^{\prime}}\, ,
\end{equation}
where $N_{\mathrm{s},0}$ is the initial photon number of the weak photon
field at the discrete signal frequency $\omega$, $N_{\mathrm{i},0}$ is the
initial idler photon number at a frequency of $2\omega_\mathrm{p}-\omega$,
and $C_{\mathrm{si},0}$ is the initial signal-idler mode correlation. Now and
in the following, it is important to distinguish between the input and the
output modes, as well as the inner degrees of freedom of the amplifier. The
input and output of the amplifier is given by a weak signal photon field at a
frequency $\omega$ with a narrow bandwidth $B_\mathrm{s}\ll\omega$. The total
photon number at the output of the amplifier within the narrow signal
bandwidth $B_\mathrm{s}$, after a reference-mode travel-time of
$t_\mathrm{r}$, can be evaluated by integrating \eqref{eqn-correlation} over
frequency
\begin{equation}
  N_{\mathrm{s},t_{\mathrm{r}}}=\int\limits_{\omega-\frac{B_\mathrm{s}}{2}}^{\omega+\frac{B_\mathrm{s}}{2}}\varexpect{\op{A}^\dagger_{\omega}\op{A}_{\omega^{\prime}}}\diff{\omega^\prime}\, .\label{eqn-photon-number-integral}
\end{equation}
Performing an integration by parts of \eqref{eqn-photon-number-integral},
considering the delta functions in the spectral densities, yields the total
photon number $N_{\mathrm{s},t_\mathrm{r}}$ at the output, given by

\begin{widetext}
  \begin{equation}
    \begin{split}
      N_{\mathrm{s},t_{\mathrm{r}}}&=\function{\bar{n}}{\omega}+\left[N_{\mathrm{s},0}-\function{\bar{n}}{\omega}\right]\cdot\function{\zeta_1}{\omega,t_\mathrm{r}}\function{\zeta_1^\ast}{\omega,t_\mathrm{r}}\e^{-\frac{\function{\gamma}{\omega}+\function{\gamma}{2\omega_\mathrm{p}-\omega}}{2}t_\mathrm{r}}+C_{\mathrm{s}\mathrm{i},0}\cdot\function{\zeta_1}{\omega,t_\mathrm{r}}\function{\zeta_2^\ast}{\omega,t_\mathrm{r}}\e^{-\frac{\function{\gamma}{\omega}+\function{\gamma}{2\omega_\mathrm{p}-\omega}}{2}t_\mathrm{r}}\\
      &\quad+\left[N_{\mathrm{i},0}+\function{\bar{n}}{\omega}+1\right]\cdot\function{\zeta_2}{\omega,t_\mathrm{r}}\function{\zeta_2^\ast}{\omega,t_\mathrm{r}}\e^{-\frac{\function{\gamma}{\omega}+\function{\gamma}{2\omega_\mathrm{p}-\omega}}{2}t_\mathrm{r}}+C_{\mathrm{s}\mathrm{i},0}^\ast\cdot\function{\zeta_2}{\omega,t_\mathrm{r}}\function{\zeta_1^\ast}{\omega,t_\mathrm{r}}\e^{-\frac{\function{\gamma}{\omega}+\function{\gamma}{2\omega_\mathrm{p}-\omega}}{2}t_\mathrm{r}}\\
      &\quad+\left[\function{\bar{n}}{\omega}+\function{\bar{n}}{2\omega_\mathrm{p}-\omega}+1\right]\cdot\function{\bar{F}}{\omega,t_\mathrm{r}}\e^{-\frac{\function{\gamma}{\omega}+\function{\gamma}{2\omega_\mathrm{p}-\omega}}{2}t_\mathrm{r}}\, ,
    \end{split}
    \label{eqn-photon-number}
  \end{equation}
  where the evolution of the independent part $\function{\bar{F}}{\omega}$ in
  the third line of \eqref{eqn-photon-number} is equal to
  \begin{equation}
    \begin{split}
      \function{\bar{F}}{\omega}&=-\frac{\abs{g}^2\abs{\rho}^2\function{\gamma}{2\omega_\mathrm{p}-\omega}^2\cdot\function{\zeta_1}{t}\function{\zeta_1^\ast}{t}}{\function{\gamma}{\omega}\function{\gamma}{2\omega_\mathrm{p}-\omega}\left[4\abs{g}^2\abs{\eta}^2+\function{\gamma}{\omega}\function{\gamma}{2\omega_\mathrm{p}-\omega}\right]-\left[\function{\gamma}{\omega}+\function{\gamma}{2\omega_\mathrm{p}-\omega}\right]^2\abs{g}^2\abs{\rho}^2}\\
      &\quad+\frac{\abs{g}^2\abs{\rho}^2\left[\function{\gamma}{\omega}\function{\gamma}{2\omega_\mathrm{p}-\omega}+\function{\gamma}{2\omega_\mathrm{p}-\omega}^2\right]\cdot\left[\function{\zeta_2}{t}\function{\zeta_2^\ast}{t}+\e^{\frac{\function{\gamma}{\omega}+\function{\gamma}{2\omega_\mathrm{p}-\omega}}{2}t}\right]}{\function{\gamma}{\omega}\function{\gamma}{2\omega_\mathrm{p}-\omega}\left[4\abs{g}^2\abs{\eta}^2+\function{\gamma}{\omega}\function{\gamma}{2\omega_\mathrm{p}-\omega}\right]-\left[\function{\gamma}{\omega}+\function{\gamma}{2\omega_\mathrm{p}-\omega}\right]^2\abs{g}^2\abs{\rho}^2}\\
      &\quad-\frac{\function{\gamma}{\omega}\function{\gamma}{2\omega_\mathrm{p}-\omega}\cdot\abs[\big]{g\rho\cdot\function{\zeta_1}{t}+\left[2g\eta+\function{\gamma}{2\omega_\mathrm{p}-\omega}\right]\cdot\function{\zeta_2}{t}}^2}{\function{\gamma}{\omega}\function{\gamma}{2\omega_\mathrm{p}-\omega}\left[4\abs{g}^2\abs{\eta}^2+\function{\gamma}{\omega}\function{\gamma}{2\omega_\mathrm{p}-\omega}\right]-\left[\function{\gamma}{\omega}+\function{\gamma}{2\omega_\mathrm{p}-\omega}\right]^2\abs{g}^2\abs{\rho}^2}\, .\label{eqn-independent-noise}
    \end{split}
  \end{equation}
\end{widetext}

The results in \eqref{eqn-photon-number} and \eqref{eqn-independent-noise}
are extending the analytic solutions from~\cite{holm_quantum_1987} (when
expressed in terms of complex exponentials). In contrast to the original
work, our solution also includes a non-zero phase-mismatch as well as
chromatic dispersion in the time domain.

\subsection{Gain and resonant phase-matching}

From~\eqref{eqn-photon-number}, we obtain the well-known gain spectrum for a
JTWPA~\cite{houde_loss_2019,reep_mesoscopic_2019} with additional exponential
damping due to the inclusion of substrate losses. The gain $G$ is given by
the factor in front of the initial signal photon number $N_{\mathrm{s},0}$ at
$t_\mathrm{r}=0$, i.e.\
\begin{equation}
  G=\function{\zeta_1}{\omega,t_{\mathrm{r}}}\function{\zeta_1^\ast}{\omega,t_{\mathrm{r}}}\e^{-\frac{\function{\gamma}{\omega}+\function{\gamma}{2\omega_{\mathrm{p}}-\omega}}{2}t_{\mathrm{r}}}\, .\label{eqn-signal-power-gain}
\end{equation}
Apart from the power gain spectrum in \eqref{eqn-signal-power-gain}, which is
determined by the parametric amplification of the signal photon field
including substrate losses~\cite{houde_loss_2019}, one can define a quantum
gain of a JTWPA as the ratio of the expected photon number at the output to
the input photon number, which is given by
\begin{equation}
  G_{\mathrm{q}}= \frac{N_{\omega,t_{\mathrm{r}}}}{N_{\omega,0}}\, .
  \label{eqn-quantum-signal-gain}
\end{equation}
In the classical limit, i.e.\ for a large number of input photons
$N_{\omega,0}$ compared to the idler field and additional noise, the
converted idler photons and thermal fluctuations can be neglected and we
recover the power signal gain, as defined in~\eqref{eqn-signal-power-gain}.

It is crucial to minimize the frequency-dependent total phase-mismatch
$\Delta{\Omega}$ in order to achieve a high parametric gain over a large
bandwidth. From~\eqref{eqn-total-phase-mismatch}, it is clear that the total
phase-mismatch $\Delta{\Omega}$ is affected by self- and
cross-phase-modulation effects, as well as chromatic
dispersion~\cite{reep_mesoscopic_2019}. Thus, proper dispersion engineering
by including resonant phase-matchers~\cite{obrien_resonant_2014} or spatial
modulations of the transmission line
structure~\cite{planat_photonic-crystal_2020} can have a huge impact on the
overall amplifier performance. Resonant phase-shifters, as depicted in blue
colour in \figref{fig-unit-cell}, can be incorporated into our model by
adjusting the dispersion relation accordingly. This is done by replacing the
ground capacitance $C^{\prime}\Delta{z}$ by a frequency dependent impedance
$\function{Z}{\omega}$~\cite{grimsmo_squeezing_2017} in each unit cell
\begin{equation}
  \function{Z}{\omega}=\left[\imu\omega C^{\prime}\Delta{z}+\frac{\imu\omega C_{\mathrm{c}}\left(1-L_{\mathrm{r}}C_{\mathrm{r}}\omega^2\right)}{1-\left(C_{\mathrm{r}}+C_{\mathrm{c}}\right)L_{\mathrm{r}}\omega^2}\right]^{-1}\, .
\end{equation}
We still use the same reference timeframe $t=t_{\mathrm{r}}$ from
Section~\ref{sec-time-frame} where, however, the dispersion factor
$\function{\Lambda_{\mathrm{RPM}}}{\omega}$ has to be adjusted according to
\begin{equation}
  \pdd{\function{t_{\mathrm{RPM}}}{\omega}}{t_{\mathrm{r}}}=\sqrt{\function{\Lambda_{\mathrm{RPM}}}{\omega}}=\sqrt{\frac{\function{\Lambda}{\omega}}{\imu\omega\function{Z}{\omega}C^{\prime}\Delta{z}}}\, .
\end{equation}

\subsection{Treatment of losses}

In microwave engineering, substrate losses are typically expressed in terms
of a loss tangent, which is defined as the ratio of the total effective
conductivity to the lossless permittivity of the
substrate~\cite{pozar_microwave_2012}
\begin{equation}
  \tan\delta=\frac{\omega\varepsilon^{\prime\prime}+\sigma}{\omega\varepsilon^{\prime}}\, ,
\end{equation}
where $\omega$ is the angular frequency, $\sigma$ is the conductivity per
unit length and
$\varepsilon=\varepsilon^{\prime}+\imu\varepsilon^{\prime\prime}$ is the
permittivity of the substrate. Assuming that all internal losses are
homogeneously distributed and only attributed to the substrate, the linear
attenuation factor $\alpha$ can be calculated from the lossless propagation
constant $k_0$ by
\begin{equation}
  \alpha=\frac{k_0\tan\delta}{2}\, .
\end{equation}
Note that this simple linear loss model does not take into account additional
losses due to e.g.\ periodic loadings along the line, or pump saturation
effects, which can however be included in a more elaborate approach for
finding the respective coupling parameters $\function{\gamma}{\omega}$. The
corresponding exponential damping factor $\function{\tilde{\Gamma}}{\omega}$
of a lossy JTWPA can hence be expressed as a function of the loss tangent
\begin{equation*}
  \function{\tilde{\Gamma}}{\omega}=\frac{1}{2}\omega\sqrt{\frac{C'\Delta{z}}{L_{\mathrm{J},0}}}Z_0\cdot\tan\delta\, ,
\end{equation*}
with the characteristic impedance $Z_0$ of the transmission line, where we
have already switched the spatial description of the attenuation factor to a
temporal description and taken into account the translation between the
reference travel time $t_{\mathrm{r}}$ and the frequency dependent
propagation time $\function{t}{\omega}$. The classical damping factor
$\function{\tilde{\Gamma}}{\omega}$ can then be related to the quantum
mechanical damping coefficients $\function{\gamma}{\omega}$ of our model by
$2\function{\tilde{\Gamma}}{\omega}\mapsto\function{\gamma}{\omega}$.

\subsection{Modeling of JTWPAs}

We solved~\eqref{eqn-quantum-signal-gain} and~\eqref{eqn-photon-number} for a
well-studied structure from the
literature~\cite{obrien_resonant_2014,reep_mesoscopic_2019} and used the
result to demonstrate the parametric gain for a large number of input
photons, see~\eqref{eqn-signal-power-gain}. The device parameters of the
JTWPA are given in~\tabref{tab-gain-spectrum}.
\begin{table}[b!]
  \caption{\label{tab-gain-spectrum}JTWPA device parameters from~\cite{obrien_resonant_2014,reep_mesoscopic_2019}.}
  \begin{ruledtabular}
    \begin{tabular}{lll}
      Parameter                           & Symbol            & Value                    \\[0.5ex] \hline
      Josephson capacitance               & $C_\mathrm{J}$    & \SI{329}{\femto\farad}   \\
      Josephson junction critical current & $I_\mathrm{c}$    & \SI{3.29}{\micro\ampere} \\
      Pump current amplitude              & $I_\mathrm{p}$    & \num{0.5}$I_\mathrm{c}$  \\
      RPM coupling capacitance            & $C_\mathrm{c}$    & \SI{10}{\femto\farad}    \\
      RPM resonator capacitance           & $C_\mathrm{r}$    & \SI{7.036}{\pico\farad}  \\
      RPM resonator inductance            & $L_\mathrm{r}$    & \SI{100}{\pico\henry}    \\
      Substrate loss tangent              & $\tan\delta$      & \num{0.0025}             \\
      Unit cell ground capacitance        & $C$               & \SI{39}{\femto\farad}    \\
      Unit cell number                    & $N_\mathrm{cell}$ & \num{2000}               \\
      Unit cell physical length           & $l_\mathrm{cell}$ & \SI{10}{\micro\meter}
    \end{tabular}
  \end{ruledtabular}
\end{table}

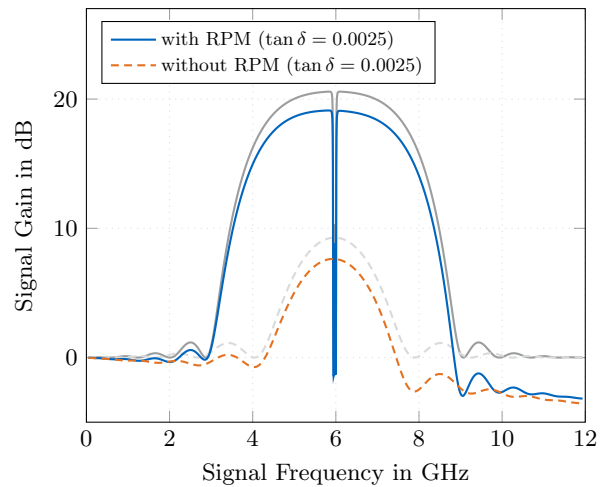
\begin{figure}[t]
  \centering
  \begin{tikzpicture}
    \sisetup{round-mode=places,round-precision=4}
    \begin{axis}[
        legend pos = north west,
        legend cell align = left,
        legend style={nodes={scale=0.8, transform shape}},
        width = 0.95\columnwidth,
        xmin = 0,
        xmax = 12e9,
        ymin = -5,
        ymax = 27,
        change x base, x SI prefix=giga,
        unit marking pre = {\text{in }},
        unit marking post = {},
        x unit = \si{\hertz},
        y unit = \si{\deci\bel},
        xlabel = {Signal Frequency},
        ylabel = {Signal Gain},
        grid = both,
        grid style = {draw=TUMGrayLight, dotted, ultra thin},
      ]
      \addplot[color=TUMGray, thick, solid, forget plot] table [y=rpm_lossless, x=f, col sep=comma] {fig-gain-spectrum-obrien_2014.csv};
      \addplot[color=TUMGrayLight, thick, densely dashed, forget plot] table [y=norpm_lossless, x=f, col sep=comma] {fig-gain-spectrum-obrien_2014.csv};
      \addplot[color=TUMBlue, thick, solid] table [y=rpm_lossy, x=f, col sep=comma] {fig-gain-spectrum-obrien_2014.csv};
      \addplot[color=TUMOrange, thick, densely dashed] table [y=norpm_lossy, x=f, col sep=comma] {fig-gain-spectrum-obrien_2014.csv};
      \legend{with RPM ($\tan\delta=\num{0.0025}$), without RPM ($\tan\delta=\num{0.0025}$)}
    \end{axis}
  \end{tikzpicture}
  \caption{\label{fig-gain-spectrum} Gain spectrum of a Josephson traveling-wave parametric amplifier including substrate losses with a loss tangent $\tan\delta=\num{0.0025}$. The parameters of the JTWPA structure are taken from the literature~\cite{obrien_resonant_2014,reep_mesoscopic_2019} and are explicitly given in~\tabref{tab-gain-spectrum}. The blue line describes the gain spectrum with resonant phase-matchers placed in every unit cell, while the dashed orange curve represents the signal gain without dispersion engineering. The gray lines represent the gain spectra for the lossless case with (solid) and without (dashed) phase-matching, respectively.}
\end{figure}

In~\figref{fig-gain-spectrum} the signal gain is given as a function of
frequency. We have included both cases with and without resonant
phase-matching according to~\cite{obrien_resonant_2014}. The literature
results~\cite{obrien_resonant_2014,reep_mesoscopic_2019} for an ideal
substrate could be reproduced by our model and are given in gray color
in~\figref{fig-gain-spectrum}. The resulting gain considering a more
realistic substrate material with a loss tangent $\num{0.0025}$ is given in
solid blue and dashed orange color, for the case with and without resonant
phase-matching, respectively.

\begin{figure}[thb]
  \centering
  \begin{tikzpicture}
    \begin{axis}[
        ymode=log,
        legend pos=north west,
        legend cell align=left,
        legend style={nodes={scale=0.8, transform shape}},
        width=0.95\columnwidth,
        xmin = 0,
        xmax = 4e-9,
        ymin = 7.5e-1,
        ymax = 225,
        change x base, x SI prefix=nano,
        unit marking pre={\text{in }},
        unit marking post={},
        x unit=\si{\second},
        xlabel={Time},
        ylabel={Signal Photon Number $N_{\omega}$},
        grid=both,
        grid style={draw=TUMGrayLight, dotted, ultra thin},
      ]
      \addplot[color=TUMGray, thick, solid, forget plot] table [y=rpm_5_lossless, x=t, col sep=comma] {fig-temporal-dynamics-obrien_2014.csv};
      \addplot[color=TUMGrayLight, thick, solid, forget plot] table [y=rpm_4_lossless, x=t, col sep=comma] {fig-temporal-dynamics-obrien_2014.csv};
      \addplot[color=TUMGray, thick, densely dashed, forget plot] table [y=norpm_5_lossless, x=t, col sep=comma] {fig-temporal-dynamics-obrien_2014.csv};
      \addplot[color=TUMGrayLight, thick, densely dashed, forget plot] table [y=norpm_4_lossless, x=t, col sep=comma] {fig-temporal-dynamics-obrien_2014.csv};
      \addplot[color=TUMBlue, thick, solid] table [y=rpm_5_lossy, x=t, col sep=comma] {fig-temporal-dynamics-obrien_2014.csv};
      \addplot[color=TUMOrange, thick, solid] table [y=rpm_4_lossy, x=t, col sep=comma] {fig-temporal-dynamics-obrien_2014.csv};
      \addplot[color=TUMBlue, thick, densely dashed] table [y=norpm_5_lossy, x=t, col sep=comma] {fig-temporal-dynamics-obrien_2014.csv};
      \addplot[color=TUMOrange, thick, densely dashed] table [y=norpm_4_lossy, x=t, col sep=comma] {fig-temporal-dynamics-obrien_2014.csv};
      \legend{with RPM ($f_\mathrm{s}=\SI{5}{\giga\hertz}$), with RPM ($f_\mathrm{s}=\SI{4}{\giga\hertz}$), without RPM ($f_\mathrm{s}=\SI{5}{\giga\hertz}$), without RPM ($f_\mathrm{s}=\SI{4}{\giga\hertz}$)}
    \end{axis}
  \end{tikzpicture}
  \caption{\label{fig-temporal-dynamics} Temporal evolution of the number of signal photons within the JTWPA from~\tabref{tab-gain-spectrum}. The blue lines describe the temporal dynamics of a signal mode with a center frequency of \SI{5}{\giga\hertz} with (solid) and without (dashed) RPM. The orange curve represents the time evolution of a signal at \SI{4}{\giga\hertz}, also with (solid) and without (dashed) RPM. The solid gray lines represent the temporal evolution for the lossless case at both frequencies, while the dashed lines correspond to the case without phase-matching.}
\end{figure}
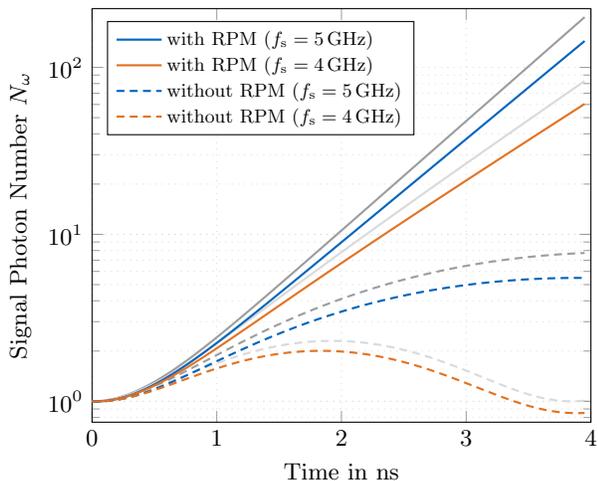

It can be seen that proper dispersion engineering can substantially enhance
the device's performance in terms of gain and bandwidth, as pointed out
in~\cite{obrien_resonant_2014}. This becomes even more clear when taking a
look at the temporal dynamics of the expected photon number in the signal
mode. Assuming a single signal photon at the input of the JTWPA, the temporal
evolution of the signal photon number is depicted
in~\figref{fig-temporal-dynamics}. The time argument on the abscissa is given
in terms of the reference timeframe from Section~\ref{sec-time-frame}. The
signal photon number at around $\SI{4}{\nano\second}$ corresponds to the
output of the amplifier after \num{2000} unit cells. With proper dispersion
engineering, exponential gain can be achieved as it can be seen from the
solid blue and orange curves in \figref{fig-temporal-dynamics}, corresponding
to signal frequencies of \SI{5}{\giga\hertz} and \SI{4}{\giga\hertz},
respectively. According to the gain spectrum in~\figref{fig-gain-spectrum}, a
parametric gain of $\approx\SI{15}{\deci\bel}$ can be achieved for a mode at
\SI{4}{\giga\hertz} with RPM. Without RPM, however, the gain attains a local
minimum at \SI{4}{\giga\hertz} and even becomes negative when dielectric
losses are taken into account. The temporal dynamics
in~\figref{fig-temporal-dynamics} reveal that without RPM, the signal mode at
\SI{4}{\giga\hertz} first experiences some gain until around
\SI{1.75}{\nano\second}. From there on, the signal mode gets depopulated due
to the large phase-mismatch of the signal and pump modes, which results in an
average photon number smaller than one at the output due to the transmission
line losses. At \SI{5}{\giga\hertz}, however, there is still some gain of
around \SI{5}{\deci\bel}, even without RPM.

\begin{figure}[t]
  \centering
  \begin{tikzpicture}
    \sisetup{round-mode=places,round-precision=4}
    \begin{axis}[
        legend pos = north west,
        legend cell align = left,
        legend columns = 3,
        legend style={nodes={scale=0.8, transform shape}},
        width = 0.95\columnwidth,
        xmin = 4e9,
        xmax = 8e9,
        ymin = -10,
        ymax = 17,
        change x base, x SI prefix=giga,
        unit marking pre = {\text{in }},
        unit marking post = {},
        x unit = \si{\hertz},
        y unit = \si{\deci\bel},
        xlabel = {Signal Frequency},
        ylabel = {Signal Gain},
        grid = both,
        grid style = {draw=TUMGrayLight, dotted, ultra thin},
      ]
      \addplot[color=TUMGray, thick, solid] table [y=experiment, x=f, col sep=comma] {fig-gain-spectrum-experiment-calibrated.csv};
      \addplot[color=TUMBlue, thick, solid] table [y=gain_dB, x=f, col sep=comma] {fig-gain-norpm-lossy_simbierowicz_2021_fit.csv};
      \addplot[color=TUMOrange, thick, solid] table [y=gain_dB, x=f, col sep=comma] {fig-gain-norpm-lossy_simbierowicz_2021.csv};
      \coordinate (insetPosition) at (rel axis cs:0.25, 0.15);
      \legend{Experiment, Theory (fit), Theory}
    \end{axis}
    \begin{axis}[
        at={(insetPosition)},
        footnotesize,
        axis background/.style={fill=white},
        width=0.25\textwidth,
        height=0.175\textwidth,
        xmin = 0,
        xmax = 12e9,
        ymin = -10,
        ymax = 15,
        change x base, x SI prefix=giga,
        unit marking pre = {\text{in }},
        unit marking post = {},
        grid = both, grid style = {draw=TUMGrayLight, dotted, ultra thin}, ]
      \addplot[color=TUMGray, thick, solid] table [y=experiment, x=f, col sep=comma] {fig-gain-spectrum-experiment-calibrated.csv};
      \addplot[color=TUMBlue, thick, solid] table [y=gain_dB, x=f, col sep=comma] {fig-gain-norpm-lossy_simbierowicz_2021_fit.csv};
      \addplot[color=TUMOrange, thick, solid] table [y=gain_dB, x=f, col sep=comma] {fig-gain-norpm-lossy_simbierowicz_2021.csv};
      \draw[very thick, densely dashed, TUMGreen] (axis cs:4e9,-9.75) rectangle (axis cs:8e9,14.75);
    \end{axis}
  \end{tikzpicture}
  \caption{\label{fig-gain-spectrum-experiment} Gain spectrum of the experimental JTWPA from~\cite{simbierowicz_characterizing_2021} from \SIrange{4}{8}{\giga\hertz}. The respective parameters for the JTWPA unit cell are given in~\tabref{tab-gain-spectrum-experiment}. The noisy gray line in the background show the measurement results for the JTWPA, extracted from \figurename~3 of~\cite{simbierowicz_characterizing_2021}. The orange line depicts the resulting gain spectrum of our model using exactly the parameters given in~\tabref{tab-gain-spectrum-experiment}. A better fit (blue line) to the experimental data was achieved by slightly increasing the unit cell capacitance to $\SI{140}{\femto\farad}$ while decreasing the pump current to $\num{0.51}I_\mathrm{c}$ to account for the neglected pump depletion. The inset shows the gain spectrum for the full \SIrange{0}{12}{\giga\hertz} frequency range.}
\end{figure}
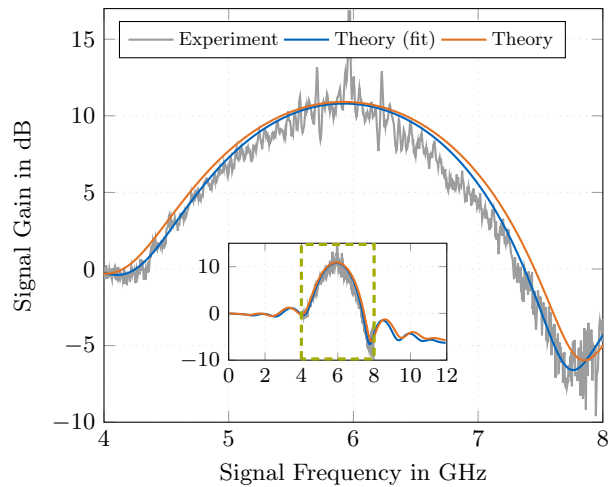

The device parameters from \cite{obrien_resonant_2014,reep_mesoscopic_2019}
have so far only been studied theoretically. We now want to assess how our
model performs for an experimentally realized structure, that has been
reported in~\cite{groenberg_side-wall_2017,simbierowicz_characterizing_2021}.
Equations \eqref{eqn-photon-number} and \eqref{eqn-signal-power-gain} are
used to predict the power gain of the signal mode, with the nonlinear
transmission line parameters from~\cite{simbierowicz_characterizing_2021}.
The respective device parameters are summarized
in~\tabref{tab-gain-spectrum-experiment}. We then compare our analytic
results with the measured gain spectrum from
\cite{simbierowicz_characterizing_2021} and demonstrate that our model is
suitable as a design tool for predicting the device performance of realistic
JTWPAs.
\begin{table}[b]
  \caption{\label{tab-gain-spectrum-experiment}JTWPA device parameters from~\cite{simbierowicz_characterizing_2021}.}
  \begin{ruledtabular}
    \begin{tabular}{lll}
      Parameter                           & Symbol              & Value                             \\[0.5ex] \hline
      Josephson plasma frequency          & $\omega_\mathrm{J}$ & $2\pi\cdot\SI{46.5}{\giga\hertz}$ \\
      Josephson junction critical current & $I_\mathrm{c}$      & $\SI{4.4}{\micro\ampere}$         \\
      Pump current amplitude              & $I_\mathrm{p}$      & $\num{0.53}I_\mathrm{c}$          \\
      Substrate loss tangent              & $\tan\delta$        & $\num{0.0025}$                    \\
      Unit cell ground capacitance        & $C$                 & $\SI{115}{\femto\farad}$          \\
      Unit cell inductance (unpumped)     & $L$                 & $\SI{312}{\pico\henry}$           \\
      Unit cell number                    & $N_\mathrm{cell}$   & $\num{1016}$                      \\
      Unit cell physical length           & $l_\mathrm{cell}$   & $\SI{26}{\micro\meter}$
    \end{tabular}
  \end{ruledtabular}
\end{table}
In this particular device, no dispersion engineering was employed. Thus, the
provided results comprise a relatively large phase-mismatch compared to the
previous resonantly phase-matched device, albeit not as large as for the case
without RPM in~\cite{obrien_resonant_2014,reep_mesoscopic_2019}. The
resulting gain spectrum with the parameters from
\tabref{tab-gain-spectrum-experiment} is given in terms of the orange curve
in~\figref{fig-gain-spectrum-experiment}. The noisy gray line was extracted
from~\cite{simbierowicz_characterizing_2021} and represents the
experimentally determined gain spectrum. Comparing both curves shows good
agreement of the predicted and measured spectra. An even better agreement was
achieved by slightly modifying the device parameters. To account for pump
depletion, we slightly decreased the pump current, assumed to be constant in
our model, from $\num{0.53}I_\mathrm{c}$ to $\num{0.51}I_\mathrm{c}$. At the
same time, we increased the unit cell ground capacitance $C$ from
$\SI{115}{\femto\farad}$ to $\SI{140}{\femto\farad}$. This results in an
almost perfect match over the entire frequency range that was covered in the
experiment (see the blue curve in \figref{fig-gain-spectrum-experiment}). The
inset plot shows the full spectrum as predicted by the analytic model, where
the dashed green rectangle marks the frequency range where experimental data
was available.

\section{\label{sec-noise}Added noise analysis}

Quantitatively, the noise figure of conventional low-noise amplifiers is due
to the resistive nature of the amplification mechanism. Superconducting
parametric amplifiers, in contrast, rely on nonlinear wave-mixing processes,
where the added noise is not limited by
dissipation~\cite{kubo_fluctuation-dissipation_1966}. Hence, the noise
performance of a JTWPA is only restricted by quantum mechanical limits on the
fluctuations of the signals involved in the wave-mixing
process~\cite{haus_quantum_1962}. The noise at the output of a linear
amplifier can be separated into contributions of the amplified input
fluctuations, and the noise added by the amplifier's internal degrees of
freedom. The quantum noise is bounded by the uncertainty principle and gives
rise to a standard quantum limit (SQL). In the following, we investigate the
added noise of JTWPAs including substrate losses and associated fluctuations,
and compare the predictions of our theory to experimental results.

Fluctuations within a quantum limited amplifiers are usually evaluated based
on the average of the field variance~\cite{haus_quantum_1962}. The quantum
limit can then be given by a single added noise number $A$, defined as the
number of energy quanta added to the input of a perfectly noiseless linear
amplifier~\cite{caves_quantum_1982,caves_quantum_2012}. In the
literature~\cite{houde_loss_2019,zhao_quantum_2021,fasolo_bimodal_2022}, this
approach has been applied to study the effect of loss asymmetries as well as
thermal and quantum noise in traveling-wave parametric amplifiers.
According to~\cite{caves_quantum_1982}, at this point, a distinction needs to
be made in terms of the actual input and output signals. In our case, we
consider a weak photon field at frequency $\omega$ with a narrow bandwidth
$B_\mathrm{s}\ll\omega$, which carries the information. This signal mode gets
amplified by four-wave mixing while traveling along the Josephson
junction-embedded nonlinear transmission line, experiencing parametric gain,
substrate losses, and noise added by the wave-mixing and the environmental
fluctuations. The mean-square fluctuations $\abs{\Delta\op{A}_\omega}^2$
within the narrow bandwidth $B_\mathrm{s}$ of the annihilation operator
$\op{A}_\omega$ of the weak photon field are given by the symmetric variance
\begin{equation}
  \abs{\Delta\op{A}_{\omega,t_\mathrm{r}}}^2=\int\limits_{\omega-\frac{B_\mathrm{s}}{2}}^{\omega+\frac{B_\mathrm{s}}{2}}\frac{1}{2}\varexpect{\op{A}_{\omega,t_\mathrm{r}}\op{A}_{\omega^\prime,t_\mathrm{r}}^\dagger+\op{A}_{\omega,t_\mathrm{r}}^\dagger\op{A}_{\omega^\prime,t_\mathrm{r}}}\diff{\omega^\prime}\, ,
\end{equation}
assuming vanishing expectation values
$\varexpect{\op{A}_{\omega,t_\mathrm{r}}}$ and
$\varexpect{\op{A}_{\omega,t_\mathrm{r}}^\dagger}$, which is the case for
thermal and number states~\cite{barnett_methods_1997}. Using the Bosonic
commutator relation
$\commutator{\op{A}_{\omega,t_\mathrm{r}}}{\op{A}_{\omega^\prime,t_\mathrm{r}}^\dagger}=\function{\delta}{\omega-\omega^\prime}$,
it holds that after a reference interaction time $t_\mathrm{r}$, the output
fluctuations are equal to
\begin{equation}
  \abs{\Delta\op{A}_{\omega,t_\mathrm{r}}}^2=\int\limits_{\omega-\frac{B_\mathrm{s}}{2}}^{\omega+\frac{B_\mathrm{s}}{2}}\varexpect{\op{A}_\omega^\dagger\op{A}_{\omega^\prime}}\diff{\omega^\prime}+\frac{1}{2}=N_{\mathrm{s},t_\mathrm{r}}+\frac{1}{2}\, .\label{eqn-output-fluctuations}
\end{equation}
Similarly, the mean-square fluctuations of the weak photon field at the input
of the amplifier are given by
$\abs{\Delta\op{A}_{\omega,0}}^2=N_{\mathrm{s},0}+1/2$. The mean-square
fluctuations at the input and output of the amplifier are given in terms of
energy quanta. Hence, when propagating through the device, the input
fluctuations experience gain, while the inner degrees of freedom within the
JTWPA add additional noise quanta~\cite{caves_quantum_1982}. This is
reflected by the simple linear gain model
\begin{equation}
  \begin{split}
    \abs{\Delta\op{A}_{\omega,t_\mathrm{r}}}^2&=G\cdot\abs{\Delta\op{A}_{\omega,0}}^2+\abs{\Delta\op{\mathcal{F}}}_\mathrm{op}^2\\
    &=G\cdot\left[\abs{\Delta\op{A}_{\omega,0}}^2+\frac{\abs{\Delta\op{\mathcal{F}}}_\mathrm{op}^2}{G}\right]\, ,
  \end{split}\label{eqn-fluctuation-gain-model}
\end{equation}
where $\abs{\Delta\op{\mathcal{F}}}_\mathrm{op}^2$ represents the noise due
to the inner degrees of freedom of the amplifier, borrowing the notation
from~\cite{caves_quantum_1982}. The added noise $A$, i.e.\ the energy quanta
added to the input due to the intrinsic quantum fluctuations of the
amplifier, is then given by $A=\abs{\Delta\op{\mathcal{F}}}_\mathrm{op}^2/G$.
Thus, we can calculate the added noise $A$, given the expressions for the
input and output fluctuations, i.e.\
\begin{equation}
  A=\frac{\abs{\Delta\op{A}_{\omega,t_\mathrm{r}}}^2}{G}-\abs{\Delta\op{A}_{\omega,0}}^2=\frac{N_{\mathrm{s},t_\mathrm{r}}+\frac{1}{2}}{G}-N_{\mathrm{s},0}-\frac{1}{2}\, .\label{eqn-added-noise-abstract}
\end{equation}
Finally, it remains to determine an expression for the gain $G$, that is
experienced by the input fluctuations in order to get the fluctuations at the
output. From \eqref{eqn-fluctuation-gain-model}, we see that the gain $G$ is
given by the multiplicative factor in front of the input fluctuations
$\abs{\Delta\op{A}_{\omega,0}}^2=N_{\mathrm{s},0}+1/2$ in the expression of
the output mean-square fluctuations
$\abs{\Delta\op{A}_{\omega,t_\mathrm{r}}}^2=N_{\mathrm{s},t_\mathrm{r}}+1/2$.
By rearranging~\eqref{eqn-photon-number}, it can be seen that
$G=\function{\zeta_1}{\omega,t_{\mathrm{r}}}\function{\zeta_1^\ast}{\omega,t_{\mathrm{r}}}\exp{\left\lbrace-\left[\function{\gamma}{\omega}+\function{\gamma}{2\omega_{\mathrm{p}}-\omega}\right]t_{\mathrm{r}}/2\right\rbrace}$,
which is exactly equal to the signal power gain
in~\eqref{eqn-signal-power-gain}, as expected.
Inserting the analytic solution for $N_{\mathrm{s},t_\mathrm{r}}$
from~\eqref{eqn-photon-number} and the expression for the power gain $G$
from~\eqref{eqn-signal-power-gain} into \eqref{eqn-added-noise-abstract}
yields the overall added noise of the parametric amplifier with substrate
loss, which is given by
\begin{widetext}
  \begin{equation}
    \begin{split}
      A&=\left[\function{\bar{n}}{\omega}+\frac{1}{2}\right]\cdot\left[\frac{1}{\function{\zeta_1}{\omega,t_\mathrm{r}}\function{\zeta_1^\ast}{\omega,t_\mathrm{r}}\e^{-\frac{\function{\gamma}{\omega}+\function{\gamma}{2\omega_\mathrm{p}-\omega}}{2}t_\mathrm{r}}}-1\right]+C_{\mathrm{s}\mathrm{i},0}\cdot\frac{\function{\zeta_2^\ast}{\omega,t_\mathrm{r}}}{\function{\zeta_1^\ast}{\omega,t_\mathrm{r}}}+C_{\mathrm{s}\mathrm{i},0}^\ast\cdot\frac{\function{\zeta_2}{\omega,t_\mathrm{r}}}{\function{\zeta_1}{\omega,t_\mathrm{r}}}\\
      &\quad+\left[N_{\mathrm{i},0}+\function{\bar{n}}{\omega}+1\right]\cdot\frac{\function{\zeta_2}{\omega,t_\mathrm{r}}\function{\zeta_2^\ast}{\omega,t_\mathrm{r}}}{\function{\zeta_1}{\omega,t_\mathrm{r}}\function{\zeta_1^\ast}{\omega,t_\mathrm{r}}}+\left[\function{\bar{n}}{\omega}+\function{\bar{n}}{2\omega_\mathrm{p}-\omega}+1\right]\cdot\frac{\function{\bar{F}}{\omega,t_\mathrm{r}}}{\function{\zeta_1}{\omega,t_\mathrm{r}}\function{\zeta_1^\ast}{\omega,t_\mathrm{r}}}\, .
    \end{split}\label{eqn-added-noise}
  \end{equation}
\end{widetext}
The first term in the second line of~\eqref{eqn-added-noise} gives the noise
contributions due to the wave mixing process with the fluctuations of the
idler mode. The last term represents fluctuations due to thermal noise from
the substrate interactions, which is particularly sensitive to the bath
temperature. A lower bound on the added noise $A$ for phase-insensitive
linear amplifiers was derived and proven in~\cite{caves_quantum_1982}, which
is given by
\begin{equation}
  A\geq\frac{1}{2}\abs*{1-\frac{1}{G}}\, ,\qquad\text{for }G\geq 1\, .\label{eqn-added-noise-limit}
\end{equation}
Thus, in the high-gain limit $G\to\infty$ we obtain the half-photon standard
quantum limit (SQL) $\lim\limits_{G\to\infty}A=1/2$, which should be a
reasonable limit within the spectral regions where one finally wants to
operates a JTWPA. As a sanity check for our model, we investigated the
lossless case where $T\to 0$ and $\gamma\to 0$. If we then let
$G=\abs{\function{\zeta_1}{\omega,t_\mathrm{r}}}^2\to\infty$, one can easily
see that~\eqref{eqn-added-noise} also arives at the SQL.

\begin{figure}[t]
  \centering
  \begin{tikzpicture}
    \pgfplotsset{
      xmin = 0,
      xmax = 12e9,
      grid = both,
      grid style = {draw=TUMGrayLight, dotted, ultra thin},
      width = 0.88\columnwidth,
      unit marking pre = {\text{in }},
      unit marking post = {},
      legend pos = north west,
      legend cell align = left,
      legend columns = 3,
      legend style={nodes={scale=0.525, transform shape}},
    }
    \begin{axis}[
        set layers = axis lines on top,
        change x base, x SI prefix=giga,
        xlabel = {Signal Frequency},
        unit marking pre = {\text{in }},
        unit marking post = {},
        x unit = \si{\hertz},
        ymin = -1,
        ymax = 2,
        ylabel = {Added Noise Photons},
        axis y line*=left,
        ytick = {-1,-0.5,0,0.5,1,1.5,2},
      ]
      \draw[TUMExtRed, thick, dashdotted, name path=sql] (axis cs:0,0.5) -- (axis cs:12e9,0.5) node[pos=0.1, fill=white, scale=0.75] {SQL};
      \draw[TUMIvory, dashdotted, name path=zero_line] (axis cs:0,0.0) -- (axis cs:12e9,0.0);
      \draw[draw=none,TUMIvory, dashdotted, name path=bottom_line] (axis cs:0,-2.5) -- (axis cs:12e9,-2.5);
      \addplot[color=TUMBlue, thick, solid] table [y=noise, x=f, col sep=comma] {fig-noise-rpm-lossy_obrien_2014.csv};
      \addplot[color=TUMOrange, thick, solid] table [y=noise, x=f, col sep=comma] {fig-noise-norpm-lossy_obrien_2014.csv};
      \addplot[name path=limit_rpm,color=TUMGray, thick, solid] table [y=limit, x=f, col sep=comma] {fig-noise-rpm-lossy_obrien_2014.csv};
      \addplot[name path=limit,color=TUMGrayLight, thick, solid] table [y=limit, x=f, col sep=comma] {fig-noise-norpm-lossy_obrien_2014.csv};
      \addplot [TUMIvory!25, forget plot] fill between [of = limit_rpm and limit];
      \addplot [TUMIvory!25, forget plot] fill between [of = limit_rpm and zero_line];
      \addplot [TUMIvory!25, forget plot] fill between [of = limit and zero_line];
      \addplot [TUMIvory!25, forget plot] fill between [of = zero_line and bottom_line];
      \draw[very thick, densely dashed, TUMGreen] (axis cs:4.5e9,0.35) rectangle (axis cs:7.5e9,0.75);
      %
      \coordinate (insetPosition) at (rel axis cs:0.1075,0.075);
      \legend{Add.\ noise (RPM), Add.\ noise, Limit (RPM), Limit}
      \addlegendimage{/pgfplots/refstyle=plt_noise_obrien_gain_rpm}\addlegendentry{Gain (RPM)}
      \addlegendimage{/pgfplots/refstyle=plt_noise_obrien_gain}\addlegendentry{Gain}
      \addlegendimage{area legend,color=TUMIvory!25,fill}
      %
    \end{axis}
    \begin{axis}[
        ymin = -30,
        ymax = 30,
        y unit = \si{\deci\bel},
        ylabel = {Signal Gain},
        axis y line*=right,
        grid = none,
        hide x axis,
      ]
      \addplot[color=TUMBlue, densely dashed] table [y=gain_dB, x=f, col sep=comma] {fig-noise-rpm-lossy_obrien_2014.csv};
      \label{plt_noise_obrien_gain_rpm}
      \addplot[color=TUMOrange, densely  dashed] table [y=gain_dB, x=f, col sep=comma] {fig-noise-norpm-lossy_obrien_2014.csv};
      \label{plt_noise_obrien_gain}
      \draw[TUMBlue,-latex] (axis cs:3e9,6.75) -- (axis cs:3.875e9,6.75);
      \draw[TUMOrange,-latex] (axis cs:5e9,6.75) -- (axis cs:5.875e9,6.75);
      %
      %
    \end{axis}
    \begin{axis}[
        at={(insetPosition)},
        footnotesize,
        axis background/.style={fill=white},
        width=0.2525\textwidth,
        height=0.1475\textwidth,
        xmin = 4.5e9,
        xmax = 7.5e9,
        ymin = 0.35,
        ymax = 0.75,
        change x base, x SI prefix=giga,
        ytick = {0.3,0.5,0.4,0.6,0.7},
        grid = both,
        grid style = {draw=TUMGrayLight, dotted, ultra thin},
        y tick label style={
            /pgf/number format/.cd,
            fixed,
            fixed zerofill,
            precision=1,
            /tikz/.cd
          }
      ]
      \draw[TUMExtRed, thick, dashdotted, name path=sql] (axis cs:4.5e9,0.5) -- (axis cs:7.5e9,0.5) node[pos=0.75, fill=white, scale=0.5] {SQL};
      \addplot[color=TUMBlue, thick, solid] table [y=noise, x=f, col sep=comma] {fig-noise-rpm-lossy_obrien_2014.csv};
      \addplot[color=TUMOrange, thick, solid] table [y=noise, x=f, col sep=comma] {fig-noise-norpm-lossy_obrien_2014.csv};
      \addplot[name path=limit_fit,color=TUMGray, thick, solid] table [y=limit, x=f, col sep=comma] {fig-noise-rpm-lossy_obrien_2014.csv};
      \addplot[name path=limit,color=TUMGrayLight, thick, solid] table [y=limit, x=f, col sep=comma] {fig-noise-norpm-lossy_obrien_2014.csv};
    \end{axis}
  \end{tikzpicture}
  \caption{\label{fig-noise-obrien}Added noise of the JTWPA from~\cite{obrien_resonant_2014,reep_mesoscopic_2019} including substrate losses with a loss tangent $\tan\delta=\num{0.0025}$ at $T=\SI{50}{\milli\kelvin}$. The blue line describes the added noise with resonant phase-matchers placed in every unit cell, while the orange curve represents the added noise without dispersion engineering. The gray lines represent the limits on the added noise, with (dark gray) and without (light gray) phase-matching. Additionally, the gain spectrum is given by the blue and orange dashed line for both cases (right axis). The inset shows a zoom-in of the added noise in the spectral range with a large gain. The gray shaded area marks the forbidden range for the case with RPM.}
\end{figure}
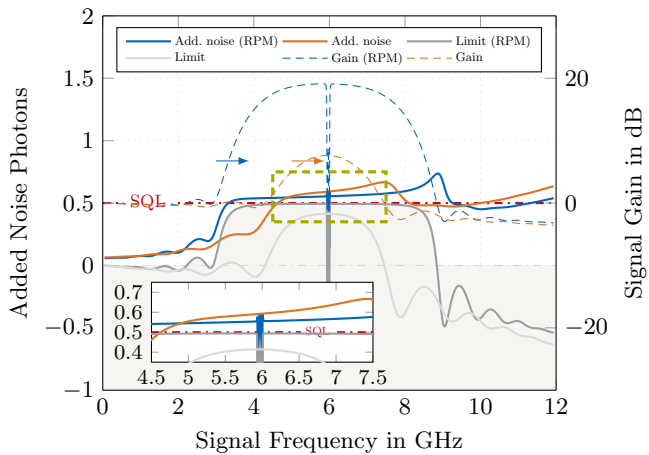
In Section~\ref{sec-analytic-solution}, we solved for the gain spectra of two
parameter sets of different JTWPAs from the literature. In the following, we
present our results on the added noise for both parameter sets, which are
computed by solving~\eqref{eqn-added-noise}. Let us first investigate the
added noise of the JTWPA
from~\cite{obrien_resonant_2014,reep_mesoscopic_2019}, with the parameters
summarized in~\tabref{tab-gain-spectrum}. For this particular parameter set,
we assume a bath temperature $T=\SI{50}{\milli\kelvin}$.
In~\figref{fig-noise-obrien}, the added noise is given as a function of
frequency. The solid orange line represents the added noise of the JTWPA
without dispersion engineering. The dashed orange line depicts the gain
spectrum for this case. The light gray line represents the minimum added
noise for the poorly phase-matched case. One can clearly see from the solid
blue curve, which depicts the added noise over frequency including RPM, that
proper dispersion engineering not only increases gain and bandwidth, but also
considerably reduces the equivalent added noise at the input of the amplifier
to around $\num{0.55}$ quanta on average, just slightly above the quantum
limit, given in dark gray color. The gain for the resonantly phase-mathced
case is given by the dashed blue line. The gray shaded region is the
forbidden area for the added noise in the dispersion engineered case, as the
added noise cannot become negative, which is due to the fact that the gain
must be larger or equal to one for the limit
in~\eqref{eqn-added-noise-limit}. If chromatic dispersion is properly
compensated for, the quantum limit is close to the SQL, as the gain becomes
relatively large, which is apparent from the inset in
\figref{fig-noise-obrien}, which shows a zoom-in of the usable bandwidth of
the amplifier. This device, as mentioned before, has however only been
studied theoretically in~\cite{obrien_resonant_2014} and
\cite{reep_mesoscopic_2019}. In order to show the value and validity of our
analytic model including losses and associated noise, we now have a closer
look at the added noise of the experimentally studied JTWPA device
from~\cite{groenberg_side-wall_2017,simbierowicz_characterizing_2021}.

\begin{figure}[t]
  \centering
  \begin{tikzpicture}
    \pgfplotsset{
      xmin = 0,
      xmax = 12e9,
      grid = both,
      grid style = {draw=TUMGrayLight, dotted, ultra thin},
      width = 0.88\columnwidth,
      unit marking pre = {\text{in }},
      unit marking post = {},
      legend pos = north west,
      legend cell align = left,
      legend columns = 1,
      legend style={nodes={scale=0.525, transform shape}},
    }
    \begin{axis}[
        set layers = axis lines on top,
        change x base, x SI prefix=giga,
        xlabel = {Signal Frequency},
        unit marking pre = {\text{in }},
        unit marking post = {},
        x unit = \si{\hertz},
        ymin = -2,
        ymax = 4,
        ylabel = {Added Noise Photons},
        axis y line*=left,
        ytick = {-2,-1,0,1,2,3,4},
      ]
      \draw[TUMExtRed, thick, dashdotted, name path=sql] (axis cs:0,0.5) -- (axis cs:12e9,0.5) node[pos=0.75, fill=white, scale=0.75] {SQL};
      \draw[TUMIvory, dashdotted, name path=zero_line] (axis cs:0,0.0) -- (axis cs:12e9,0.0);
      \draw[draw=none,TUMIvory, dashdotted, name path=bottom_line] (axis cs:0,-2.5) -- (axis cs:12e9,-2.5);
      \addplot[color=TUMBlue, thick, solid] table [y=noise, x=f, col sep=comma] {fig-noise-norpm-lossy_simbierowicz_2021_fit.csv};
      \addplot[color=TUMOrange, thick, solid] table [y=noise, x=f, col sep=comma] {fig-noise-norpm-lossy_simbierowicz_2021.csv};
      \addplot[name path=limit_fit,color=TUMGray, thick, solid] table [y=limit, x=f, col sep=comma] {fig-noise-norpm-lossy_simbierowicz_2021_fit.csv};
      \addplot[name path=limit,color=TUMGrayLight, thick, solid] table [y=limit, x=f, col sep=comma] {fig-noise-norpm-lossy_simbierowicz_2021.csv};
      \addplot [TUMIvory!25, forget plot] fill between [of = limit_fit and limit];
      \addplot [TUMIvory!25, forget plot] fill between [of = limit_fit and zero_line];
      \addplot [TUMIvory!25, forget plot] fill between [of = limit and zero_line];
      \addplot [TUMIvory!25, forget plot] fill between [of = zero_line and bottom_line];
      \draw[very thick, densely dashed, TUMGreen] (axis cs:5e9,0.2) rectangle (axis cs:7e9,1.2);
      \node[TUMGreen, above, scale=0.75] at (axis cs:6e9,1.2) {$+\num{0.5}$};
      \coordinate (insetPosition) at (rel axis cs:0.175,0.075);
      \legend{Add.\ noise (fit), Add.\ noise, Limit (fit), Limit}
      \addlegendimage{/pgfplots/refstyle=plt_noise_simbierowicz_gain_fit}\addlegendentry{Gain (fit)}
      \addlegendimage{/pgfplots/refstyle=plt_noise_simbierowicz_gain}\addlegendentry{Gain}
      \addlegendimage{no markers,TUMBlueLight,thick}
      \addlegendentry{Experiment}
      \addlegendimage{area legend,color=TUMIvory!25,fill}
      \addlegendentry{Forbidden}
    \end{axis}
    \begin{axis}[
        ymin = -8.9375,
        ymax = 12.5,
        y unit = \si{\deci\bel},
        ylabel = {Signal Gain},
        axis y line*=right,
        grid = none,
        hide x axis,
      ]
      \addplot[color=TUMBlue, densely dashed] table [y=gain_dB, x=f, col sep=comma] {fig-noise-norpm-lossy_simbierowicz_2021_fit.csv};
      \label{plt_noise_simbierowicz_gain_fit}
      \addplot[color=TUMOrange, densely  dashed] table [y=gain_dB, x=f, col sep=comma] {fig-noise-norpm-lossy_simbierowicz_2021.csv};
      \label{plt_noise_simbierowicz_gain}
      \draw[TUMBlue,-latex] (axis cs:4.625e9,7.5) -- (axis cs:5.5e9,7.5);
      \draw[TUMOrange,-latex] (axis cs:4.625e9,6.5) -- (axis cs:5.5e9,6.5);
      %
      %
    \end{axis}
    \begin{axis}[
        at={(insetPosition)},
        footnotesize,
        axis background/.style={fill=white},
        width=0.225\textwidth,
        height=0.1525\textwidth,
        xmin = 5e9,
        xmax = 7e9,
        ymin = 0.7,
        ymax = 1.7,
        change x base, x SI prefix=giga,
        ylabel = {Noise},
        ytick = {0.7,1,1.3,1.6},
        grid = both,
        grid style = {draw=TUMGrayLight, dotted, ultra thin},
        y tick label style={
            /pgf/number format/.cd,
            fixed,
            fixed zerofill,
            precision=1,
            /tikz/.cd
          }
      ]
      \fill[TUMBlueLighter, opacity=0.5] (axis cs:5e9,1.1) rectangle (axis cs:7e9,1.6);
      \draw[TUMBlueLight, thick] (axis cs:5e9,1.3) -- (axis cs:7e9,1.3);
      \draw[TUMExtRed, thick, dashdotted, name path=sql] (axis cs:5e9,1) -- (axis cs:7e9,1) node[pos=0.75, fill=white, scale=0.5] {SQL};
      \addplot[color=TUMBlue, thick, solid] table [y expr=\thisrow{noise}+0.5, x=f, col sep=comma] {fig-noise-norpm-lossy_simbierowicz_2021_fit.csv};
      \addplot[color=TUMOrange, thick, solid] table [y expr=\thisrow{noise}+0.5, x=f, col sep=comma] {fig-noise-norpm-lossy_simbierowicz_2021.csv};
      \addplot[name path=limit_fit,color=TUMGray, thick, solid] table [y expr=\thisrow{limit}+0.5, x=f, col sep=comma] {fig-noise-norpm-lossy_simbierowicz_2021_fit.csv};
      \addplot[name path=limit,color=TUMGrayLight, thick, solid] table [y expr=\thisrow{limit}+0.5, x=f, col sep=comma] {fig-noise-norpm-lossy_simbierowicz_2021.csv};
    \end{axis}
  \end{tikzpicture}
  \caption{\label{fig-noise-simbierowicz} Added noise of the experimental JTWPA from~\cite{simbierowicz_characterizing_2021} for a temperature $T=\SI{20}{\milli\kelvin}$. The orange line depicts the added noise calculated with our model using exactly the parameters given in~\tabref{tab-gain-spectrum-experiment}. The blue line corresponds to the modified parameter set (see \figref{fig-gain-spectrum-experiment}). The gray lines mark the minimum added noise for the original and modified parameter sets, respectively. The filled area in gray color represents the forbidden area. Additionally, the respective gain spectra are given by the thin dashed orange and blue lines. The inset shows the total equivalent noise at the input of the JTWPA in the usable spectral range within the highlighted green area. Excellent matching with the experimentally determined input noise of $\num{1.3}_{-0.2}^{+0.3}$ from~\cite{simbierowicz_characterizing_2021} is obtained.}
\end{figure}
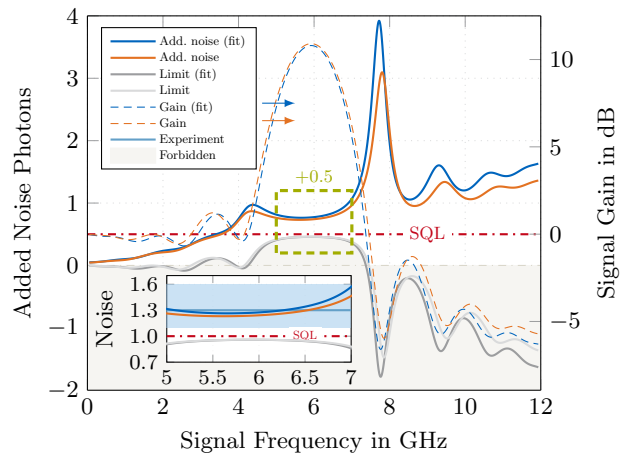

Evaluating~\eqref{eqn-added-noise} with the parameter set given
in~\tabref{tab-gain-spectrum-experiment} yields the results depicted in
\figref{fig-noise-simbierowicz}. The associated experiments have been carried
out at a temperature of $T=\SI{20}{\milli\kelvin}$, which we also used for
our modeling. In Section~\ref{sec-analytic-solution}, we introduced a
modified parameter set with slight variations on the line capacitance and the
critical current in order to get a better fit for the gain spectrum to the
experimental results. Thus, we also depict both results in
\figref{fig-noise-simbierowicz}, where the solid and dashed orange lines
correspond to the added noise and gain spectrum for the original parameter
set given in \tabref{tab-gain-spectrum-experiment}, and the blue solid and
dashed lines represent the added noise and gain spectrum for the modified
parameters. Also here, we show the minimum added noise according
to~\eqref{eqn-added-noise-limit} in light and dark gray colors for the
original and modified parameter sets, respectively. Compared to the previous
device from~\cite{obrien_resonant_2014}, one can see that the added noise is
somewhat higher, in the range of around $\num{0.8}$ quanta on average. In the
experiment in~\cite{simbierowicz_characterizing_2021}, the noise temperature
at the input of an amplifier cascade including the JTWPA was measured with
the so-called Y-factor method~\cite{pozar_microwave_2012}. The intrinsic
noise of the JTWPA was then calulated by subtracting different signal paths,
one that included the JTWPA and one that was short-circuited. The noise
temperature translates into the total variance at the input of the amplifier,
including the half-quantum in~\eqref{eqn-output-fluctuations} for
$t_\mathrm{r}=0$. Thus, in the inset in \figref{fig-noise-simbierowicz}, we
added this half-quantum to be able to compare the resulting input
fluctuations with the experimental results of $\num{1.3}_{-0.2}^{+0.3}$. The
experimental input photon variance is given in light blue, with the
uncertainty range highlighted as light blue area. The predictions of our
theory on the added input fluctuations are precisely within the uncertainty
of the experimental observation, as apparent from the inset of
\figref{fig-noise-simbierowicz}.

\section{\label{sec-conclusion}Conclusion}

In this paper, we introduced a rather complete quantum circuit model for a
Josephson traveling-wave parametric amplifier operating in the four-wave
mixing mode, including self- and cross-phase-modulation, phase mismatching
due to chromatic dispersion, as well as losses and associated fluctuations
due to the imperfect substrate isolation. In contrast to existing
literature~\cite{obrien_resonant_2014,grimsmo_squeezing_2017,houde_loss_2019,planat_photonic-crystal_2020}
where the amplification is investigated over the interaction length in terms
of a spatial description, we presented the temporal evolution of the mode
operators based on the Heisenberg equations in the scattering limit (i.e.,
assuming perfect impedance matching). An attempt for such a temporal
formulation was given in \cite{reep_mesoscopic_2019}, however, without proper
accounting for dispersion. We included dispersion by introducing a reference
timeframe with associated dispersion factors in the respective signal and
idler modes. From the system and bath dynamics, we formulated coupled mode
temporal evolution equations for the creation and annihilation operators of
the weak idler and signal photon fields. Even for the general case, including
non-zero dispersion and phase matching as well as substrate losses and
associated fluctuations, we found an analytic solution for the signal
annihilation operator, assuming a strong classical monochromatic pump tone.
Losses are here included by means of a loss tangent that is associated with
the bath interaction strength. Using this analytic solution, we found an
expression for the temporal evolution of the photon number in the signal
mode. Based on this, we were able to predict the gain spectrum of the
amplifier, which was validated against results from the literature for the
lossless case. A direct comparison of the gain spectrum predicted by our
model to results of an experimental device yielded very good agreement over
the full frequency range. Finally we derived and evaluated the noise added to
the input variance of JTWPAs. Also here, our predictions were found to be in
excellent agreement with measured results from the literature.

\begin{acknowledgments}
  The research is part of the Munich Quantum Valley, which is supported by the Bavarian state government with funds from the Hightech Agenda Bayern Plus.
\end{acknowledgments}

\bibliographystyle{apsrev4-2}
\bibliography{references}

\end{document}